\documentclass[journal]{IEEEtran}
\IEEEoverridecommandlockouts
\usepackage{array,multirow,makecell}
\usepackage[T1]{fontenc}
\usepackage{multirow}
\usepackage{graphicx}
\usepackage{xcolor}
\usepackage{booktabs}
\usepackage{subcaption}
\usepackage{cite}
\usepackage{amsmath}
\usepackage{url}
\usepackage{algpseudocode}
\usepackage{enumitem}
\usepackage{listings}
\usepackage[utf8x]{inputenc}
\usepackage{tabularray}
\usepackage{empheq}
\usepackage[most]{tcolorbox}
\tcbuselibrary{listings}
\tcbuselibrary{skins} 
\usepackage{longtable}

\usepackage{algorithm}
\usepackage{colortbl}
\usepackage{pgfplots} 
\usepackage[utf8x]{inputenc} 
\usepackage{longtable}
\usepackage{textcomp}
\usepackage{algorithm}
\usepackage{colortbl}
\usepackage{pgfplots} 
\usepackage{amssymb}
\usepackage{pifont}
\usetikzlibrary{patterns}
\usepackage{tikz}
\usepackage{xcolor}
\usetikzlibrary{arrows,shapes,positioning,shadows,trees}
\usepackage{rotating, adjustbox}
\usepackage{pgfplots}
\usepackage{hyperref}
\usepackage{fancyhdr}
\newcommand{\numc}[1]{\tikz[baseline=(char.base)]{\node[shape=circle,draw,inner sep=1.1pt] (char) {{\small {#1}}};}}

\makeatletter 

\newcommand{\linebreakand}{%
  \end{@IEEEauthorhalign}
  \hfill\mbox{}\par
  \mbox{}\hfill\begin{@IEEEauthorhalign}
}

\title{SecureFalcon: Are We There Yet in Automated Software Vulnerability Detection with LLMs?}

\author{
\IEEEauthorblockN{Mohamed~Amine~Ferrag\IEEEauthorrefmark{1}\IEEEauthorrefmark{5}, Ammar Battah\IEEEauthorrefmark{2}, Norbert Tihanyi\IEEEauthorrefmark{2}, Ridhi Jain\IEEEauthorrefmark{2},  Diana Maimu\c t\IEEEauthorrefmark{2}, \\ Fatima Alwahedi\IEEEauthorrefmark{2}, Thierry Lestable\IEEEauthorrefmark{2}, Narinderjit Singh Thandi\IEEEauthorrefmark{2}, \\  Abdechakour Mechri\IEEEauthorrefmark{2}, Merouane Debbah\IEEEauthorrefmark{3}, and  Lucas C. Cordeiro\IEEEauthorrefmark{4}}
 \\\IEEEauthorblockA{\IEEEauthorrefmark{1}Guelma University, Algeria}
 \\\IEEEauthorblockA{\IEEEauthorrefmark{2}Technology Innovation Institute, UAE}
\\\IEEEauthorblockA{\IEEEauthorrefmark{3}Khalifa University of Science and Technology, UAE}
\\\IEEEauthorblockA{\IEEEauthorrefmark{4}University of Manchester, UK}
\\\IEEEauthorblockA{\IEEEauthorrefmark{5}Corresponding author: ferrag.mohamedamine@univ-guelma.dz}
 }

\definecolor{codeblack}{rgb}{0,0.6,0}
\definecolor{codegray}{rgb}{0.5,0.5,0.5}
\definecolor{codepurple}{rgb}{0.58,0,0.82}
\definecolor{backcolour}{rgb}{0.9,0.9,0.9}
\definecolor{highlightcolor}{RGB}{255,106,106} 

\lstdefinestyle{mystyle}{
    language=C,
    backgroundcolor=\color{backcolour},
    commentstyle=\color{codeblack},
    keywordstyle=\color{blue},
    numberstyle=\tiny\color{codegray},
    stringstyle=\color{codepurple},
    basicstyle=\ttfamily\small,
    breakatwhitespace=false,
    breaklines=true,
    captionpos=b,
    keepspaces=true,
    numbers=left,
    numbersep=5pt,
    showspaces=false,
    showstringspaces=false,
    showtabs=false,
    tabsize=2,
    escapeinside={(*@}{@*)}, 
    moredelim=**[is][\sethlcolor{highlightcolor}]{@}{@},
}

\begin{document}
\maketitle
\begin{abstract}
Software vulnerabilities can cause numerous problems, including crashes, data loss, and security breaches. These issues greatly compromise quality and can negatively impact the market adoption of software applications and systems. Traditional bug-fixing methods, such as static analysis, often produce false positives. While bounded model checking, a form of Formal Verification (FV), can provide more accurate outcomes compared to static analyzers, it demands substantial resources and significantly hinders developer productivity. Can Machine Learning (ML) achieve accuracy comparable to FV methods and be used in popular instant code completion frameworks in near real-time? In this paper, we introduce \texttt{SecureFalcon}, an innovative model architecture with only 121 million parameters derived from the Falcon-40B model and explicitly tailored for classifying software vulnerabilities. To achieve the best performance, we trained our model using two datasets, namely the FormAI dataset and the FalconVulnDB. The FalconVulnDB is a combination of recent public datasets, namely the SySeVR framework, Draper VDISC, Bigvul, Diversevul, SARD Juliet, and ReVeal datasets. These datasets contain the top 25 most dangerous software weaknesses, such as CWE-119, CWE-120, CWE-476, CWE-122, CWE-190, CWE-121, CWE-78, CWE-787, CWE-20, and CWE-762. \texttt{SecureFalcon} achieves 94\% accuracy in binary classification and up to 92\% in multiclassification, with instant CPU inference times. It outperforms existing models such as BERT, RoBERTa, CodeBERT, and traditional ML algorithms, promising to push the boundaries of software vulnerability detection and instant code completion frameworks.
\end{abstract}


\section{Introduction} 
\label{sec:intro}

As we are undoubtedly passing through a digital age, technology affects every aspect of our lives~\cite{solove2004digital}. In such a context, vulnerability detection tools are essential safeguards in our continuously evolving digital landscape~\cite{abdel2023advanced}. With the emergence of new technologies, the palette of cyber threats becomes wider and more sophisticated in terms of employed techniques. Vulnerability detection tools assist in scanning, probing, and inspecting software, identifying weaknesses that could serve as entry points for potential malicious users or attackers. While integral to software security measures, vulnerability detection tools encounter inherent limitations in their scope and efficiency. Such tools typically depend on known patterns and signatures derived from synthetic datasets, making them less effective for detecting bugs in real-world software~\cite{li2024evaluating}. Furthermore, the datasets used by these tools are often too small~\cite{sun2024llm4vuln} or have a skewed distribution of vulnerable programs~\cite{jain2023code}, hampering the accuracy of static analyzer tools. Although the advancements in deep learning (DL) seem promising for vulnerability detection~\cite{Li2018vuldeepecker, Ziems2021secvulndeep, duan2019vulsniper, hanif2022vulberta}, their accuracy heavily relies on the data quality as well~\cite{jain2023code, chakraborty2021deep}. Most of the popular datasets are either fully synthetic~\cite{sard, juliet}, non-compilable~\cite{devign, bigvul, draper}, or have an unfair distribution of vulnerable vs non-vulnerable code~\cite{chakraborty2021deep, bigvul, diversevul}. Moreover, the labels associated with these datasets are subject to the detection method used. For instance, manual labeling is affected by human errors, whereas using static analysis tools to label the data can result in high false positives~\cite{habib2018many, park2016battles}. While the DL-based approaches are often employed for quick inference, they may compromise the model's accuracy in detecting bugs. Formal Verification (FV) approaches such as Model Checking (MC)  are preferred for verification in safety-critical systems~\cite{Hartonas-GarmhausenCCCG00}. Even though MC provides safety assurances for software systems~\cite{heitmeyer1998using, lahtinen2012model}, they are often expensive, even for a small piece of code. Bounded Model Checking (BMC)~\cite{bmc,MenezesAFLMSSBGTKC24} improves the traditional MC techniques by restricting the exploration to a predefined bound or depth. To prove safety in BMC for programs, we must compute the completeness threshold (CT), which can be smaller than or equal to the maximum number of loop iterations occurring in the program. Although BMC offers performance improvement to some extent, it is still expensive.  

\begin{figure}[htbp] 
\centering
\includegraphics[width=0.4\textwidth]{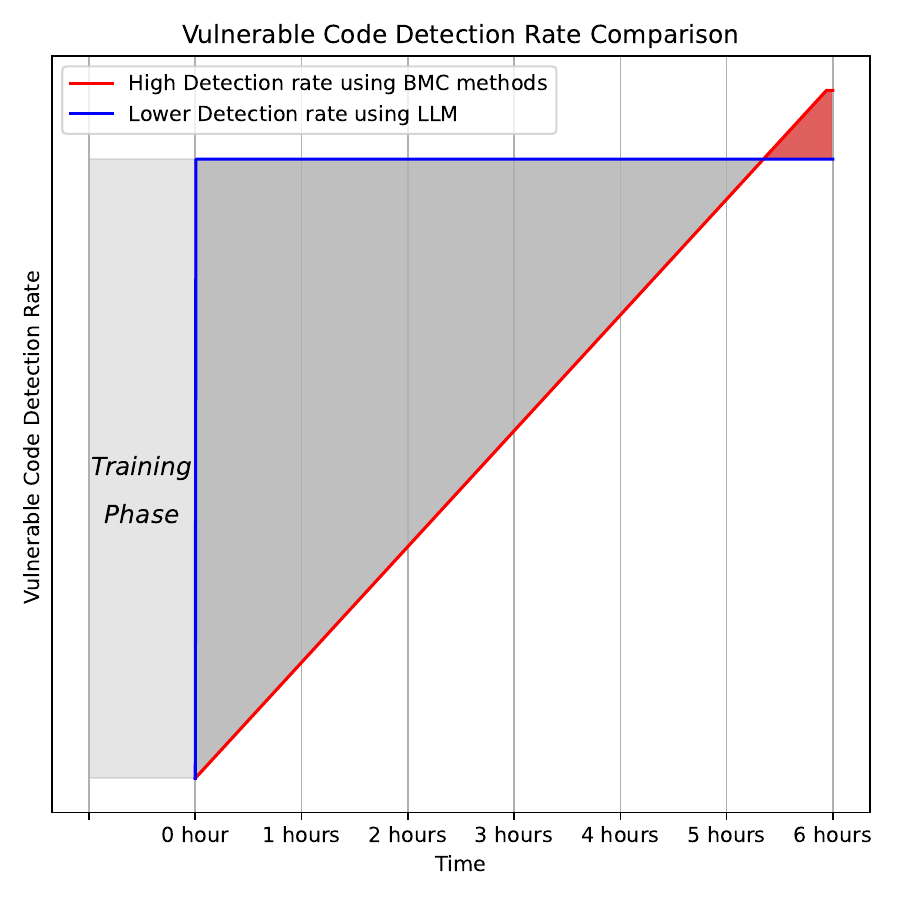}
\caption{Vulnerable Code Detection Rate (BMC vs LLMs).}
\label{fig:security_level}
\end{figure}

Code completion tools are gaining popularity in software engineering, where rapid inference is essential. For example, tools such as GitHub Copilot~\footnote{\url{https://github.com/features/copilot/}}  and Amazon Code Whisperer~\footnote{\url{https://aws.amazon.com/codewhisperer/}} suggest code snippets based on contextual analysis and training data, which, according to recent studies, can also introduce vulnerabilities~\cite{nathan_nehorai_analyzing_2024}. This raises a critical question: \textit{Can we develop a model that detects vulnerabilities efficiently without the lengthy processing times associated with BMC methods while still maintaining high accuracy?} This is essentially a trade-off between accuracy and speed. BMC methods can achieve high accuracy by eliminating false positives with counterexamples and providing stack traces. Despite this, verifying the entire state space of even simple programs can take hours. This issue is illustrated in Figure~\ref{fig:security_level}, where the red line represents the BMC method achieving high accuracy over several hours -- an impractical timeframe for real-time code completion tools. Conversely, the blue line represents a Large Language Model (LLM), which, while not reaching the same accuracy as BMC methods, still predicts high reliability and can quickly identify software vulnerabilities. 

 The red-highlighted area represents the gap that can only be closed through rigorous mathematical formal verification~\cite{DBLP:reference/mc/2018}. However, overall, formal program verification is undecidable~\cite{turing1936computable,sipser2012introduction}. We cannot devise a computational method to decide whether an arbitrary program is error-free. Much progress has been made in program verification. Nowadays, verification methods create abstractions to determine whether the program is error-free. Still, for an arbitrary program that includes unbounded memory usage, we cannot give $100$\% certainty that it is error-free due to the well-known \textit{halting problem}~\cite{turing1936computable}. However, to achieve high accuracy detection rate for vulnerable code almost instantly with a fine-tuned model would be a significant achievement. We aim to minimize this gap (red-highlighted area) as much as possible and dramatically reduce vulnerability detection times, making it practical for real-time code completion frameworks. While it is unrealistic to expect LLMs to match the results of BMC verification tools at this stage, there is promise in using ML techniques to swiftly identify the most vulnerable programs. Integrating a robust BMC tool with a pre-trained LLM can enhance the pace of vulnerability detection, benefiting the overall software development process. The success of a vulnerability detection model greatly depends on the quality and relevance of the dataset it uses. Thus, a dataset that provides a balanced distribution of vulnerable and non-vulnerable programs, mirrors real-world coding, and is labeled with a reliable method can enhance the model's accuracy. In contrast to most available C/C++ vulnerability datasets, the FormAI dataset~\cite{tihanyi2023formai} \numc{1} has high labeling accuracy as it is labeled by the Efficient SMT-based Context-Bounded Model Checker (ESBMC)~\cite{MenezesAFLMSSBGTKC24}, \numc{2} contains compilable programs, \numc{3} evenly distributes vulnerable and neutral programs, and \numc{4} is generated by LLM trained on real-world open source software, thus, mimicking the prominent developer errors and making it a highly suitable dataset for training a vulnerability detection model. The original contributions of our work are summarized as follows:
\begin{itemize}

\item We introduce \texttt{SecureFalcon}, a lightweight and innovative LLM model with only $121$ million parameters, built on the foundational 40B-parameter \textit{FalconLLM} architecture. This model offers significant enhancements and optimizations specifically tailored for security analysis. 
\texttt{SecureFalcon} was enhanced through fine-tuning using the FormAI dataset, a specialized collection designed to accurately classify vulnerabilities in C code samples. This process was supported by using the ESBMC tool, enhancing the model's detection capabilities.
\item To address the limitations of formal verification tools, which are highly effective at detecting memory-related issues such as buffer overflows (CWE-121, CWE-122) and array bound violations (CWE-129)  but are unable to identify vulnerabilities like OS command injection (CWE-78) or SQL injection (CWE-89), we created an aggregated dataset named \textit{FalconVulnDB}. This dataset integrates resources from the SySeVR framework, Draper VDISC, Bigvul, Diversevul, SARD Juliet, and ReVeal datasets to enhance \texttt{SecureFalcon}'s training with a comprehensive collection of public datasets.

These resources collectively cover examples of the top $25$ most critical software weaknesses identified by the Common Weakness Enumeration (CWE), including but not limited to CWE-119, CWE-120, CWE-476, CWE-122, CWE-190, CWE-121, CWE-78, CWE-787, CWE-20, and CWE-762. The \textit{FalconVulnDB} dataset was used to compensate for the lack of real project data in the FormAI dataset.

\item In terms of performance, \texttt{SecureFalcon} showcases exceptional proficiency in binary classification, achieving an accuracy rate of $94$\% in identifying vulnerabilities in C/C++ code. Moreover, the model maintains a robust accuracy rate of $92$\% across diverse code samples in multi-classification, underscoring its effectiveness and reliability in vulnerability detection. With these results, we outperformed traditional ML algorithms like KNN, LR, NB, SVM, RRF, DT, and LDA by 11\% (with RF achieving 81\%), and existing LLM models like BERT, CodeBERT, and RoBERTa by 4\% (with CodeBERT achieving 88\%). This advancement promises to push the boundaries of software vulnerability detection and instant code completion frameworks.

\end{itemize}

The rest of this paper is structured as follows. Section~\ref{sec:motivation} outlines the main motivation behind our research. Section~\ref{sec:related} overviews the research literature relevant to software vulnerability detection. Section~\ref{sec:finetune} discusses the methodology and approach employed to create \texttt{SecureFalcon}. Section~\ref{sec:evaluation} presents the experimental setup and the evaluation of \texttt{SecureFalcon}'s performance. Finally, we conclude with Section \ref{sec:conc} summarizing the key concepts, findings, and contributions with their corresponding implications.

\section{Motivation}
\label{sec:motivation}

As software development accelerates, so does the complexity of codebases, making identifying vulnerabilities a paramount concern. A tool that expeditiously pinpoints vulnerable code offers a proactive defense against potential exploits, minimizing the window of susceptibility and enhancing overall software security~\cite{hanif2021rise}. Rapid vulnerability detection not only safeguards against cyber threats but also instills confidence in software reliability, ensuring the delivery of secure and trustworthy applications within tight development schedules. Bounded Model Checking (BMC) is a formal verification technique used in computer science and software engineering to verify the correctness of critical hardware and software components within a finite number of steps. The BMC technique extracts the program code, the basis for generating a control-flow graph (CFG)~\cite{Aho:2006:CPT:1177220}. In this CFG, each node represents either a deterministic or non-deterministic assignment or a conditional statement. The next step involves converting the CFG into a Static Single Assignment (SSA) form and transforming it into a State Transition System (STS). The resultant STS can subsequently be converted into an SMT (Satisfiability Modulo Theories) formula that can be understood by an SMT solver, such as CVC5~\cite{DBLP:conf/tacas/BarbosaBBKLMMMN22}, Bitwuzla~\cite{DBLP:conf/cav/NiemetzP23} or  Z3~\cite{10.1007/978-3-540-78800-3_24}. SMT solver tools can ascertain whether a counterexample exists for certain properties within a specified bound $k$. Formally written, given a program $\mathcal{P}$, consider its finite state transition system, $\mathcal{TS}=\left(S, R, I\right)$. Here, $S$ is the set of states, $R \subseteq S \times S$ represents the set of transitions, and $(s_n, \cdots ,s_m) \in I \subseteq S$ represents the set of initial states. A state $s \in S$ includes the program counter value, \textit{pc}, and program variables. The initial state $s_1$ assigns the starting program location, and each transition $T=(s_i,s_{i+1}) \in R$ has a logical formula describing the constraints between states. In BMC, we define properties with logical formulas: $\phi(s)$ for safety/security properties and $\psi(s)$ for program termination. Notably, termination and error are exclusive: $\phi(s) \wedge \psi(s)$ is always unsatisfiable. A state is a deadlock if $T(s_i, s_{i+1}) \vee \phi(s)$ is unsatisfiable. The BMC problem, $\Delta_k$  can be expressed as:

\begin{equation}
\label{eq:bmc}
  \Delta_k = I(s_1) \wedge \bigwedge^{k-1}_{i=1} T(s_i, s_{i+1}) \wedge
 \bigvee^{k}_{i=1} \neg \phi(s_i).
 \end{equation}

 \noindent where $I$ is the initial states set, and $T(s_i, s_{i+1})$ the transition relation of $\mathcal{TS}$. This formula captures $\mathcal{TS}$ executions of length $k$, and if it is satisfied, a state violates $\phi$ within the $k$ limit. A counterexample, or trace, for a violated $\phi$ is a sequence of states $s_1, \ldots, s_k$. If Equation (\ref{eq:bmc}) is unsatisfiable, no error state exists within $k$ steps, indicating no vulnerabilities in the program up to the $k$ bound. Checking property violations using BMC techniques can be time-consuming even for relatively small programs due to lengthy loops that require unwinding or intricate function calls~\cite{MenezesAFLMSSBGTKC24}. Consider Listing~\ref{lst:example_ao} where an arithmetic overflow occurs due to the nature of the nested loops and the accumulation of values in the \textit{sum} variable. The code consists of two nested loops iterating from $0$ to $999,999$. Within each iteration, the product of \textit{i} * \textit{j} is added to the sum variable. Since the \textit{sum} variable accumulates these products (\textit{sum += i * j;}), the  \textit{sum} eventually surpasses the maximum value that an int can represent. When this happens, it causes arithmetic overflow, which means the result is beyond the range represented by the data type, resulting in an incorrect value being stored in the \textit{sum} variable.
 
\begin{tcolorbox}[title={Arithemtic overflow example }, coltitle=white, colbacktitle=black, colback=gray!20]
\begin{lstlisting}[language=C, caption={Arithmetic overflow on line number 7.}, label={lst:example_ao}]
#include <stdio.h>
int main() {
    int i, j;
    int sum = 0;
    for (i = 0; i < 1000000; ++i) {
        for (j = 0; j < 1000000; ++j) {
            (*@\textcolor{highlightcolor}{\textbf{sum += i * j;}}@*)
        }}
    printf("Sum:%ld\n", sum);
    return 0;
}
\end{lstlisting}
\end{tcolorbox}

Verifying this program with BMC can take hours due to nested loops requiring careful unwinding. For example, applying the Efficient SMT-based Context-Bounded Model Checker (ESBMC) took more than seven hours using $k$-induction. In contrast, a suitably trained ML algorithm, such as a Large Language Model (LLM), can quickly pinpoint such issues. LLMs have several advantages over traditional formal methods like BMC: they can learn from vast amounts of code and vulnerability patterns, enabling them to generalize and identify errors efficiently. LLMs are also capable of handling diverse codebases and detecting vulnerabilities without the extensive manual setup required for BMC. This example underscores the necessity for a compact model that can swiftly identify common errors in C programs, enabling quicker detection across various scenarios. Additionally, LLMs can be integrated into code completion frameworks with near-instant inference time, providing real-time feedback to developers and significantly accelerating the development process. This makes LLMs a superior choice over BMC for many practical applications in software detection and vulnerability analysis.

\section{Related Work}
\label{sec:related}

In recent research, LLMs have been employed for source code summarization~\cite{summarizationLLM, tarassow2023potential}, which can aid in vulnerability detection. Further, contrastive learning techniques have been applied to LLMs to improve vulnerability detection accuracy~\cite{cheng2022path}. Similarly, transfer learning has been widely used in vulnerability detection and repair with LLMs~\cite{chen2022neural, yin2020apply}. Researchers have significantly improved vulnerability detection performance by pre-training models on large code corpora and fine-tuning them on vulnerability-specific datasets. Neural machine translation techniques employed to translate code snippets into natural language aid in vulnerability detection~\cite{mahbub2023nmt, ito2023feature-nmt}. By translating code into human-readable descriptions, models can help developers understand code snippets' behavior and potential risks. SySeVR~\cite{li2021sysevr} uses DL to detect slice-level vulnerability by preserving semantic and syntactic knowledge about the vulnerabilities. Similarly, VulDeePecker~\cite{Li2018vuldeepecker} also extracts program slices to detect vulnerabilities. However, as these models are trained on semi-synthetic datasets, they fail to detect real-world vulnerabilities~\cite{chakraborty2021deep}. Devign~\cite{zhou2019devign} uses a Graph Neural Network (GNN) trained on manually labeled datasets for vulnerability identification. The Devigen dataset is constructed from the Linux kernel, QEMU, Wireshark, and FFmpeg, which generate around 58,000 graphs. However, the dataset only contains non-compilable functions. Several works have also used transformers' vulnerability detection due to their remarkable capabilities in understanding and processing natural language and code structures~\cite{feng2020codebert, liu2019roberta, hanif2022vulberta}. Their attention mechanisms enable them to capture complex relationships and patterns within text and programming languages. LLMs such as GPT-3\cite{brown2020language} have been explored for code security analysis and vulnerability detection. Researchers have used GPT to generate vulnerable code snippets and identify potential security flaws. CodeBERT~\cite{feng2020codebert}, a pre-trained language model, has been applied to vulnerability detection in source code. By fine-tuning the model on labeled vulnerability data, researchers have achieved promising results in identifying vulnerabilities such as SQL injection, cross-site scripting (XSS), and buffer overflow~\cite{codebertsql, singh2022cyber}. Transformer-based models, like BERT~\cite{devlin2018bert} and RoBERTa~\cite{liu2019roberta}, have also been utilized for vulnerability detection in various software artifacts. Despite a recent surge in the application of DL for vulnerability detection~\cite{kim2017codeclone, Li2018vuldeepecker, VulPecker2016, li2021sysevr}, a comprehensive solution with high confidence remains elusive~\cite{chakraborty2021deep,zeng2020software}. Most DL approaches suffer from four major issues: \numc{1} inadequate model, \numc{2} learning irrelevant features, \numc{3} data duplication, \numc{4} data imbalance~\cite{chakraborty2021deep}. 

To address these challenges, we employed a modern transformer model, the 40B parameter \textit{FalconLLM}, which aids in comprehending semantic dependencies through extensive training (addressing problem \numc{1}). Falcon's architecture has demonstrated superior performance to GPT-3, achieving impressive results while utilizing only a portion of the training compute budget and requiring less computation at inference time.  We fine-tuned the transformer model, leveraging the obtained understanding of natural language and learning directly from the source code (addressing problem \numc{2}). Further, we carefully pre-process the data to ensure no duplications or irrelevant features and minimize the class imbalance (addressing problems \numc{3} and \numc{4}). We utilize only a configured portion of the Falcon-40B model for a light and compact model that fits the assigned task. Such a design choice stems from language models~\cite{kaplan2020scaling} and thorough experimentation, pushing us to take a modest approach to the scale of the model to combat overfitting concerns. We fine-tune the model on C/C++ code samples to be able to differentiate between vulnerable and non-vulnerable samples. The final model, which we named \texttt{SecureFalcon}, consists of only 121 million parameters.

\section{Model Architecture} 
\label{sec:finetune}

\subsection{\textit{FalconLLM} Model} 

The FalconLLM40B \cite{falcon40b} is one of the best performing open-source models\footnote{https://huggingface.co/spaces/HuggingFaceH4/open\_llm\_leaderboard - 10 July 2023} that underwent an extensive training procedure on 384 GPUs (A100 40GB). The model's training procedure incorporated a 3D parallelism strategy that involved tensor parallelism of 8 (TP=8), pipeline parallelism of 4 (PP=4), and data parallelism of 12 (DP=12). This approach was used in conjunction with ZeRO (Zero Redundancy Optimizer) to enhance the efficiency of the training procedure.

The training hyperparameters used for \textsc{FalconLLM40B} were specifically selected to optimize the model's learning process. The precision of the model was set to bfloat16 to balance computational efficiency and numerical precision. The \textit{AdamW}!\cite{loshchilov2017decoupled} optimizer was chosen for its proven ability to achieve good results in less time. The learning rate was set at 1.85e-4 during the warm-up phase involving 4 billion tokens, followed by a cosine decay to 1.85e-5, which allows the model to converge more efficiently. The weight decay was set at 1e-1 to prevent overfitting, while Z-loss was set at 1e-4 to minimize the discrepancy between the model's predictions and the true values. Finally, the batch size was fixed at 1152 with a 100 billion token ramp-up to maximize computational throughput and stabilize the learning process.

\subsection{\textit{SecureFalcon} Model Architecture}

\begin{figure}[htbp] 
\centering
\includegraphics[width=0.50\textwidth]{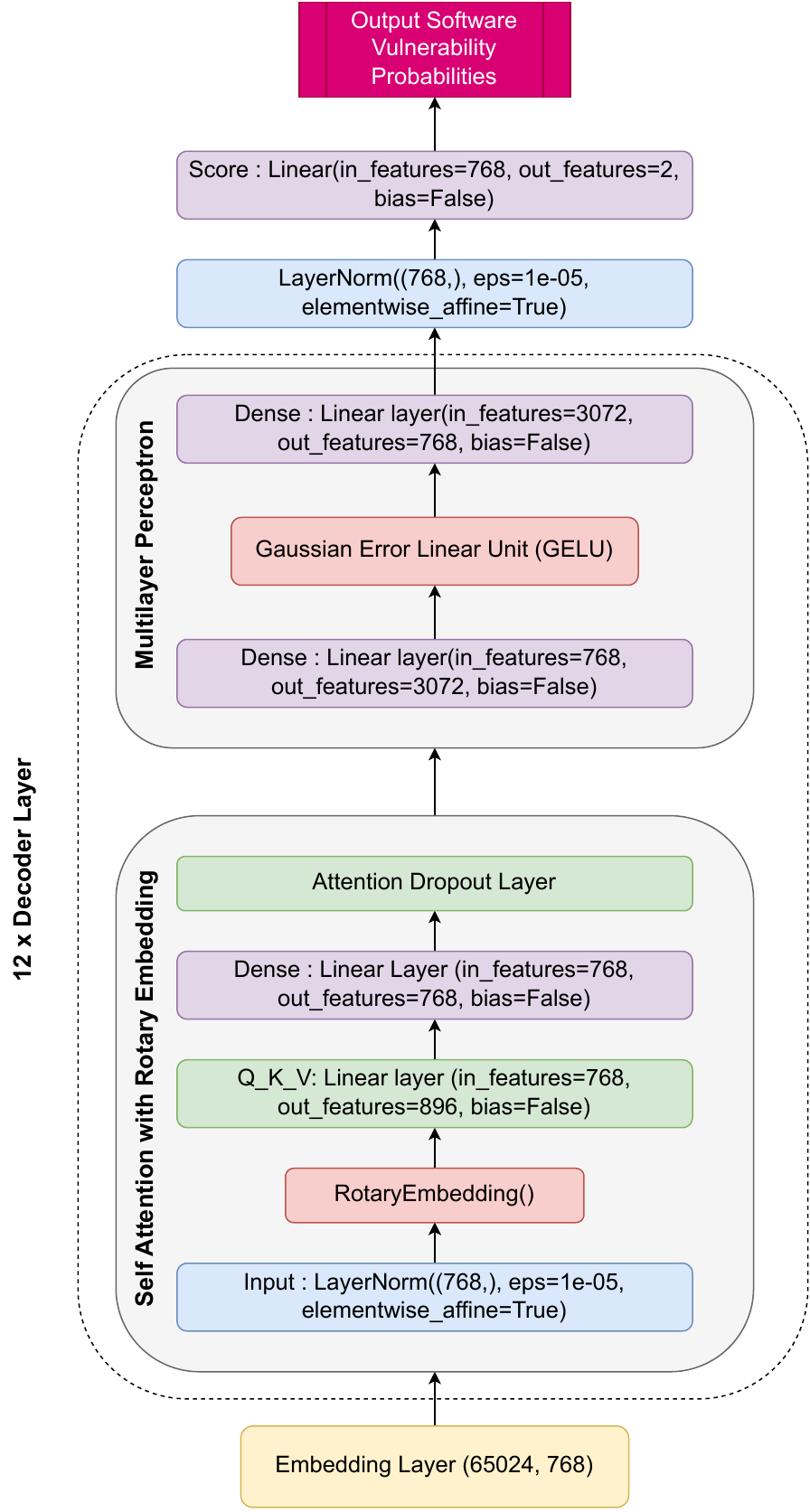}
\caption{\textit{SecureFalcon} model architecture.}
\label{fig:architecture}
\end{figure}

The \textit{SecureFalcon} model, derived from the 40B parameter FalconLLM, includes components shown in Fig. \ref{fig:architecture}, with \texttt{out\_features=12} for multi-class classification and \texttt{out\_features=2} for binary classification.

 The architecture comprises four primary components: \textit{Word Embeddings}, \textit{Encoder Layers}, \textit{Final Layer Normalization}, and the \textit{Scoring Layer}. 

\subsubsection{\textbf{Word Embeddings}} Word Embeddings serve as the initial transformation layer in the language model. This layer transforms discrete words into dense, continuous vectors, which encapsulate semantic and syntactic information of the words~\cite{mikolov2013distributed}. Their dimension is $768$, and the model is trained with a vocabulary size $65024$.

Let's denote the \(i\)-th word in an input sequence as \(input[i]\). The corresponding word embedding, \(e_i\), is a row vector obtained from the embedding matrix \(E\), which can be expressed as:

\begin{equation}
   e_{i} = E(input)[i, :]
\end{equation}


Here, \(E\) represents the embedding matrix with dimensions~\textit{vocabulary size} \(\times\) \textit{embedding dimension}, which in this case is \(65024 \times 768\). Each row in \(E\) corresponds to the vector representation of a word in the vocabulary.

Therefore, an input sequence \(input\) of length \(n\) is transformed into a sequence of word vectors \(e\) of dimension \(n \times 768\). This transformation can be depicted as:
   
\begin{equation}
    e = [e_{1}, e_{2}, \ldots, e_{n}], \text{where} \, e_{i} = E(input)[i, :]
\end{equation}
   
These word embeddings, \(e_i\), are fed into subsequent layers in the model. They encapsulate rich information about the semantics and syntactic roles of the words and their context, providing a dense representation that assists in better understanding and generating language.

\subsubsection{\textbf{Decoder Layers}} These constitute the main portion of the language model and comprise four stacked transformer layers \cite{vaswani2017attention}. Each layer includes the following components:

\begin{itemize}
    \item \textit{Layer Normalization}: This regularization technique standardizes the inputs across the feature dimension, making the model more stable and faster to train \cite{ba2016layer}. The parameters for this layer are 768-dimensional vectors. It's computed using the formula:
   
    \begin{equation}
        \hat{x} = \frac{x - \mu}{\sigma}
    \end{equation}
   
    where \(x\) is the input, \(\mu\) is the mean, \(\sigma\) is the standard deviation, and \(\hat{x}\) is the normalized output. This operation is applied to each feature independently.

   \item \textit{Self-Attention with Rotary Position Embedding (RoPE)}: Recent advancements in position encoding have shown its effectiveness within the Transformer architecture, offering valuable supervision for dependency modeling between elements at different sequence positions. In the pursuit of integrating positional information into the learning process of Transformer-based language models, a novel method, termed Rotary Position Embedding (RoPE), has been proposed by Su \textit{et al.} \cite{su2021roformer}. RoPE uniquely encodes the absolute position using a rotation matrix while incorporating explicit relative position dependency in the self-attention formulation. RoPE's implementation involves applying the rotation to the query (Q) and key (K) vectors:
   
\begin{equation}
Q', K' = \text{{rotate}}(Q, K, \text{{RoPE}})
\end{equation}

Where \(\text{{rotate}}\) is a function that applies RoPE to the vectors. The self-attention function is subsequently defined as:

\begin{equation}
\text{{Att}}(Q', K', V) = \text{{softmax}}\left(\frac{Q'K'^T}{\sqrt{d_k}}\right)V
\end{equation}

Where \(d_k\) denotes the dimension of the key vectors. The softmax function ensures that the aggregate weight of different values equals 1. Division by \(\sqrt{d_k}\) is a scaling factor to maintain gradient stability during optimization. RoPE exhibits several beneficial properties, such as sequence length flexibility, a decay in inter-token dependency with increasing relative distances, and the capability of endowing linear self-attention with relative position encoding.

   \item \textit{MLP (Multilayer Perceptron)}: The MLP is a type of neural network that comprises a minimum of three layers of nodes: an input layer, one or more hidden layers, and an output layer \cite{rumelhart1986learning}. Each layer is fully connected to the subsequent layer, signifying that every node in one layer is connected with every node in the following layer. In this case, the MLP includes an input layer. This hidden layer uses a Gaussian Error Linear Unit (GELU) activation function \cite{hendrycks2016gaussian} to introduce non-linearity and an output layer. This non-linearity allows the model to capture and learn complex patterns in the data. The MLP can be represented as:

\begin{equation}
    MLP(x) = Lin_{\text{out}}(GELU(Lin_{\text{hidden}}(x)))
\end{equation}

In this equation, \(Lin_{\text{hidden}}\) is the linear transformation corresponding to the hidden layer, \(GELU\) represents the Gaussian Error Linear Unit activation function, and \(Lin_{\text{out}}\) denotes the linear transformation of the output layer. The variable \(x\) represents the input to the MLP.

\end{itemize}

Each of the above components is essential to the functionality of the decoder layer. The layer normalization stabilizes the inputs to the self-attention and MLP components, the self-attention module allows the model to consider different parts of the input when generating each word, and the MLP provides the additional representative capacity to the model.

\subsubsection{\textbf{Final Layer Normalization}} The output from the last decoder layer undergoes an additional layer normalization operation to standardize the outputs before the final linear transformation and softmax operation. This operation follows the same mathematical principles as the layer normalization in the decoder layers. The parameters for this layer are also 768-dimensional vectors. This layer maintains a consistent distribution of activations and gradients across the network, improving model performance. In the context of language models, this helps preserve the quality of the generated text. The operation of the final layer normalization can be represented as:

\begin{equation}
    \hat{y} = \frac{y - \mu}{\sigma}
\end{equation}
   
Where \(y\) is the input to the final layer normalization, \(\mu\) is the mean, \(\sigma\) is the standard deviation, and \(\hat{y}\) is the normalized output. Similar to the layer normalization in the decoder layers, this operation is applied to each feature independently. After the final layer normalization, the 768-dimensional vectors are passed into the final linear layer and softmax function to generate the output probabilities for each word in the vocabulary.

\subsubsection{\textbf{Scoring layer}} The scoring stage involves a linear layer that generates vulnerability scores for the input software code \cite{bishop2006pattern}. This layer is engineered to transform the normalized decoder output of 768 dimensions into a 2-dimensional vector, aligning with the vulnerability classes ("vulnerable" and "not vulnerable"). Similarly, the output features are expanded to 12 dimensions for multiclass classification. 

Let's represent the decoder output, which has a dimension of 768, as \(d\). In a binary classification scenario, we denote the weight matrix as \(W\) with dimensions 768x2, and for multiclass classification, as \(W\) with dimensions 768x12. Additionally, we denote the bias vector as \(b\). Then, the score can be computed with the linear transformation as:

\begin{equation}
    \text{{Score}} = W^T d + b
\end{equation}
   
Where \(W^T d\) represents the matrix multiplication of the transpose of the weight matrix \(W\) and the decoder output vector \(d\). The score vector is then passed through a sigmoid function to convert the scores into probabilities. The sigmoid function can be defined as:
   
\begin{equation}
    P(\text{{class}}_i) = \frac{1}{1 + e^{-\text{{Score}}_i}}
\end{equation}
   
Where \(P(\text{{class}}_i)\) is the probability of the i-th class, and \(\text{{Score}}_i\) is the score for the i-th class. The final output represents the model's prediction of the vulnerability status of the input code.

The model outputs a score for each category of vulnerabilities. The class with the highest score is considered the model's prediction. Being fine-tuned from the \textit{FalconLLM} model, this architecture has proven effective for software vulnerability detection by capturing the complex syntax and semantics of programming languages, which will be demonstrated in the Experimental Evaluation Section.

\section{Datasets for Fine-tuning \texttt{SecureFalcon}}
To develop a robust model, the choice of dataset is essential. During the training phase, we utilized two datasets: the FormAI dataset~\cite{tihanyi2023formai} and FalconVulnDB, an aggregated dataset we created for fine-tuning \texttt{SecureFalcon}, compiled from all relevant datasets found in the literature.

\subsubsection{FormAI dataset}

\textit{SecureFalcon} uses the FormAI dataset\cite{tihanyi2023formai} for fine-tuning the model. The FormAI dataset includes $112,000$ compilable C code snippets created using the GPT-3.5-turbo model through a dynamic zero-shot prompting method. Since this GPT model from OpenAI\footnote{\url{https://openai.com/}} is trained using real-world open-source software repositories, the code it generates closely mimics real-world code behavior. The produced C samples are subsequently verified using ESBMC~\cite{gadelha2018esbmc}.

ESBMC is set with a verification timeout of 30 seconds per sample. According to the 2023 SV-COMP results\footnote{https://sv-comp.sosy-lab.org/2023/results/results-verified/quantilePlot-Overall.svg}, this verifier has successfully addressed the most significant number of verification tasks. The generated C samples are classified into three categories: verification successful, verification unknown, and verification failed. The verification failed class is further divided into specific vulnerability types: arithmetic overflow, buffer overflow, array bounds violation, NULL pointer dereference, division by zero, and others. The dataset provides insights into the distribution of vulnerabilities and serves as a valuable resource for vulnerability analysis and testing.

In the FormAI dataset, the classification of C programs based on vulnerability reveals that out of a total of $112,000$ programs, the verification process was performed on $106,139$ programs. Among these, $57,389$ unique programs were classified as vulnerable, resulting in $197,800$ vulnerabilities (vulnerable functions). Table \ref{tab:datadis} provides an overview of the number of samples of the FormAI dataset in two classes, `NOT VULNERABLE' (Class 0) and `VULNERABLE' (Class 1), before and after a data pre-processing stage.

\begin{table}[ht!]
\centering
\caption{Data Distribution FormAI dataset.} \label{tab:datadis}
\begin{tabular}{c|c|cc|cc}
\hline
\multirow{2}{*}{Class} & \multirow{2}{*}{Samples} & \multicolumn{2}{c|}{Before   pre-processing}      & \multicolumn{2}{c}{After   pre-processing}       \\ \cline{3-6} 
                       &                                   & \multicolumn{1}{c|}{Training} & Testing & \multicolumn{1}{c|}{Training } & Testing \\ \hline
0       & 45275                             & \multicolumn{1}{c|}{40747}         & 4528         & \multicolumn{1}{c|}{40745}         & 4528         \\
1             & 197800                            & \multicolumn{1}{c|}{178020}        & 19780        & \multicolumn{1}{c|}{55374}         & 15533        \\ \hline
\end{tabular}\\
\label{tab:FormAI}
\vspace{0.3em}
0: NOT VULNERABLE, 1: VULNERABLE 
\end{table}

Initially, the `NOT VULNERABLE' class had 40,747 training and $4,528$ testing samples, out of a total of $45,275$, with the figures remaining almost the same post-pre-processing. The `VULNERABLE' class starts with significantly more samples, $178,020$ in training, and $19,780$ in testing out of $197,800$. However, training samples reduce to $55,374$ after pre-processing, and testing samples decrease to $15,533$. Table \ref{tab:descriptive_stats} presents the key statistical measurements of the FormAI dataset. The dataset comprises $243,075$ observations in total. The average or mean value of these observations is $271.69$, with a standard deviation (Std) of 162.25, indicating the spread or variability of the dataset. The dataset's minimum (Min) value is $9$, while the maximum (Max) value observed is $2,059$. The dataset's quartile distribution is also presented, with the 25th percentile ($25$\%) at $160$, the median or the 50th percentile ($50$\%) at $235$, and the 75th percentile ($75$\%) at $343$. These figures help describe the dataset's central tendency, dispersion, and distribution shape.

\begin{table}[ht]
\centering
\caption{Statistics of FormAI dataset’s distribution.}
\label{tab:descriptive_stats}
\begin{tabular}{lr}
\hline
\textbf{Statistic} & \textbf{Value} \\
\hline
Count & 243,075 \\
Mean & 271.69 \\
Std & 162.25 \\
Min & 9 \\
25\% & 160 \\
50\% & 235 \\
75\% & 343 \\
Max & 2,059 \\
\hline
\end{tabular}
\end{table}

In addition, the vulnerabilities in the dataset are associated with registered Common Weakness Enumeration (CWE) identifiers, where each vulnerability class may map to multiple CWEs. 
The dataset comprises nine categories that address the most common vulnerabilities across 42 unique CWEs, including an "others" category containing various other instances with fewer occurrences.

\begin{enumerate}[label=\textcircled{\scriptsize\arabic*}, align=left, itemsep=0pt]
\small
\item Arithmetic overflow 
\item Buffer overflow on \texttt{scanf()/fscanf()}
\item Array bounds violated 
\item Dereference failure: NULL pointer
\item Dereference failure: forgotten memory
\item Dereference failure: invalid pointer
\item Dereference failure: array bounds violated
\item Division by zero
\item Other vulnerabilities
\end{enumerate}

Each category is associated with specific CWE numbers that capture the weaknesses leading to those vulnerabilities. For example, arithmetic overflow is associated with CWE-190, CWE-191, CWE-754, CWE-680, and CWE-681. The dictionary of the CWE-IDs associated with vulnerability types is maintained by the MITRE Corporation\footnote{\url{https://cwe.mitre.org/}}.

The pre-processing steps performed on the FormAI dataset are crucial for preparing the data before feeding it into a \textit{FalconLLM} model. The pre-processing includes removing header information and cleaning the text by removing HTML tags, links, and email addresses. These steps help standardize the text data and eliminate noise or irrelevant information hindering the subsequent analysis. In addition, we calculate the number of words in each text entry and add a new column for the word count. This new column calculates the maximum length of an input sequence that the tokenizer will accept. We convert the categorical `label' column to a numerical representation using label encoding. Generally, this pre-processing stage guarantees that the dataset is sanitized, standardized, and prepared for the fine-tuning process of \textit{FalconLLM} using the FormAI dataset. Figure~\ref{fig:dist_graph} presents the top 10 most frequent vulnerability categories in the FormAI dataset after the pre-processing. Specifically, we find that~\textit{'buffer overflow on scanf'} is the most prevalent vulnerability, recorded 29,843 times. This is followed by \textit{'dereference failure: array bounds violated'} and~\textit{'dereference failure: NULL pointer'} with 6,187 and 5,436 instances, respectively. Other notable vulnerabilities include~\textit{'dereference failure: invalid pointer'} and several types of~\textit{'arithmetic overflow'} errors, such as on~\textit{'sub'},~\textit{'add'},~\textit{'floating-point ieee\_mul'},~\textit{'floating-point ieee\_div'}, and~\textit{'mul'}, with counts ranging from 2,959 to 580.~\textit{'Buffer overflow on fscanf'} also emerges as a significant vulnerability with 625 instances.

\begin{center}
	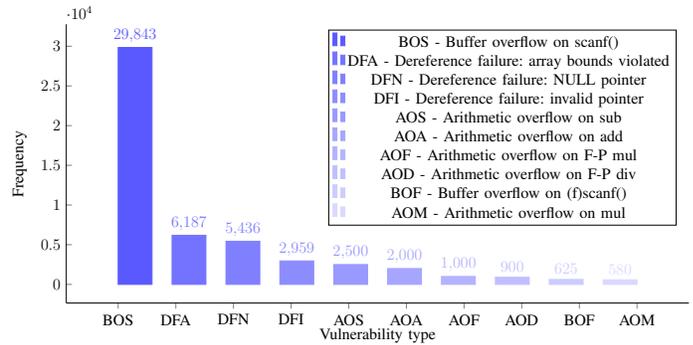
\begin{figure}[t] 
    {\large
		\begin{tikzpicture}[scale=0.5]
		\begin{axis}[
			ybar=-1cm,
			axis x line*=bottom,
			axis y line*=left,
			height=9cm, width=\textwidth,
			bar width=0.9cm,
			ylabel={Frequency},
                xlabel={Vulnerability type},
			symbolic x coords={BOS,DFA,DFN,DFI,AOS,AOA,AOF,AOD,BOF,AOM},
			x tick label style={rotate=0, anchor=north, align=left},
			nodes near coords,
			nodes near coords align={vertical}          
			]
			
			\addplot[blue!65,fill] coordinates {(BOS,29843)};
			\addplot[blue!55,fill] coordinates {(DFA,6187)};
			\addplot[blue!50,fill] coordinates {(DFN,5436)};
			\addplot[blue!45,fill] coordinates {(DFI,2959)};
			\addplot[blue!40,fill] coordinates {(AOS,2500)};
			\addplot[blue!35,fill] coordinates {(AOA,2000)};
			\addplot[blue!30,fill] coordinates {(AOF,1000)};
                \addplot[blue!25,fill] coordinates {(AOD,900)};
			\addplot[blue!20,fill] coordinates {(BOF,625)};
			\addplot[blue!15,fill] coordinates {(AOM,580)};    

        \legend{BOS - Buffer overflow on scanf(), DFA - Dereference failure: array bounds violated, DFN - Dereference failure: NULL pointer, DFI - Dereference failure: invalid pointer, AOS - Arithmetic overflow on sub, AOA - Arithmetic overflow on add, AOF - Arithmetic overflow on F-P mul, AOD - Arithmetic overflow on F-P div, BOF - Buffer overflow on (f)scanf(), AOM - Arithmetic overflow on mul}
        
		\end{axis}

		\end{tikzpicture}   
}
        \caption{Top 10 most frequent vulnerabilities categories in the FormAI dataset.}
        \label{fig:dist_graph}
	\end{figure}
 \end{center}

\subsubsection{\textcolor{black}{Limitations of AI-Generated Training Data}}

\textcolor{black}{While including the FormAI dataset allowed us to scale our training corpus with synthetically generated samples rapidly, we recognize that this dataset alone may not fully capture the complexities of real-world code. AI-generated code often lacks the historical context, unconventional patterns, and domain-specific styles in human-written codebases. To mitigate this limitation, we integrated the SySeVR framework, Draper VDISC, Bigvul, Diversevul, SARD Juliet, and ReVeal datasets into a consolidated resource called \textit{FalconVulnDB}. This aggregated dataset encompasses examples of the top 25 most critical software weaknesses (CWE), such as CWE-119, CWE-120, CWE-476, CWE-122, CWE-190, CWE-121, CWE-78, CWE-787, CWE-20, and CWE-762, drawn from public, human-curated sources.}

\textcolor{black}{By leveraging \textit{FalconVulnDB}, we aim to compensate for the scarcity of genuine project data in the FormAI corpus. Unlike AI-generated code, these publicly available datasets capture various coding habits, legacy structures, and context-driven vulnerabilities. The combination of AI-synthesized and human-sourced data fosters a more balanced training environment, encouraging our model to learn from the syntactic consistency of machine-generated samples and the gritty realism of authentic software repositories.}

\subsubsection{FalconVulnDB Dataset}

We also fine-tuned \texttt{SecureFalcon} using datasets cited in the literature, specifically: the SySeVR framework~\cite{li2021sysevr}, Draper VDISC~\cite{draper}, Bigvul~\cite{bigvul}, Diversevul~\cite{diversevul}, SARD Juliet~\cite{juliet}, andReVeal~\cite{chakraborty2021deep}. Each dataset consisted of varying elements: the first included 1,591 programs, the second held 1.27 million functions, the third contained over 264,000 functions, the fourth had more than 348,000 functions, the fifth comprised over 64,000 test cases, and the sixth featured more than 22,000 functions. Except for data from Juliet, the data is scraped from open-source projects. Test cases in Juliet are vulnerable by design and have their associated patch with them. Furthermore, we obfuscate the function and variable names as they were explicit and would affect the model's training. As for other datasets, limited pre-processing was done on them, such as function extraction, dealing with line breaks, removing comments, CWE mapping, and unifying their features. While several datasets have CWE mapping, some are missing CWE and have a CVE instead. In the case of ReVeal, it has no CWEs and is just labeled as vulnerable or not. At the same time, the Draper dataset is the only multi-labeled dataset. 

\begin{table}[ht!]
\centering
\caption{Data Distribution of FalconVulnDB.} \label{tab:datadisAD}
\begin{tabular}{c|c|cc|cc}
\hline
\multirow{2}{*}{Class} & \multirow{2}{*}{Samples} & \multicolumn{2}{c|}{Before   pre-processing}      & \multicolumn{2}{c}{After   pre-processing}       \\ \cline{3-6} 
                       &                                   & \multicolumn{1}{c|}{Training} & Testing & \multicolumn{1}{c|}{Training } & Testing \\ \hline
0       & 1,611,613                             & \multicolumn{1}{c|}{1,289,217}         & 322,396         & \multicolumn{1}{c|}{497,116}         & 124,679         \\
1             &     139,616                       & \multicolumn{1}{c|}{111,766}        & 27,850        & \multicolumn{1}{c|}{116,404}         & 28,702        \\ \hline
\end{tabular}\\
\vspace{0.3em}
0: NOT VULNERABLE, 1: VULNERABLE 
\end{table}
The amalgamation of the dataset resulted in the outcome presented in Table \ref{tab:datadisAD}. The dataset initially contained 1.7 million samples. However, there is an overlap between the different datasets, which we remove after the pre-processing. We employ more than 750,000 samples to train the model, with $20$\% being a test set. The following CWEs \footnote{https://cwe.mitre.org/top25/archive/2023/2023\_kev\_list.html} are included in FalconVulnDB.
\begin{tcolorbox}[colback=gray!10]
\scriptsize
\begin{itemize}
    \item \textbf{CWE-20: Improper Input Validation} - Occurs when software does not validate or improperly validates input, affecting a program's control or data flow. This can lead to unauthorized access, denial of service, or privilege escalation.
    
    \item \textbf{CWE-78: OS Command Injection} - 
    An application allows the execution of arbitrary OS commands due to inadequate input validation, which can result in a complete system takeover.
    
    \item \textbf{CWE-119: Improper Restriction of Operations within the Bounds of a Memory Buffer} -  A buffer overflow occurs when a program operates on more data than the size of its memory buffer. It can allow arbitrary code execution, control flow alteration, or system crash.
    
    \item \textbf{CWE-120: Buffer Copy without Checking Size of Input ('Classic Buffer Overflow')}  - A specific instance of buffer overflow caused by buffer copy operations without adequate size checks of the input.
    
    \item \textbf{CWE-121: Stack-based Buffer Overflow}  - Occurs in stack memory, potentially leading to arbitrary code execution or manipulation of program execution flow by overwriting critical data.
    
    \item \textbf{CWE-122: Heap-based Buffer Overflow} Similar to stack-based but occurs in heap memory, leading to data corruption or unexpected behavior through manipulated pointers.
    
    \item \textbf{CWE-190: Integer Overflow or Wraparound}  - Happens when an integer operation produces a value too large to be held by the integer type, causing the value to wrap and create unintended values, leading to errors or vulnerabilities.
    
    \item \textbf{CWE-476: NULL Pointer Dereference}  - It occurs when a program dereferences a pointer, which it expects to be valid but is NULL, leading to crashes or code execution.
    
    \item \textbf{CWE-762: Mismatched Memory Management Routines}  - Arises when memory is allocated and deallocated with different routines, potentially leading to heap corruption or crashes.
    
    \item \textbf{CWE-787: Out-of-bounds Write}  - It happens when software writes data outside the intended buffer boundaries, leading to data corruption, crashes, or code execution vulnerabilities.
\end{itemize}
\end{tcolorbox}
The most common vulnerability categories in FalconVulnDB are displayed in Figure~\ref{fig:dist_graph_Agg}.
\begin{center}
	\begin{figure}[ht] 
    {\large
		\begin{tikzpicture}[scale=0.45]
		\begin{axis}[
			ybar=-1cm,
			axis x line*=bottom,
			axis y line*=left,
			height=9cm, width=\textwidth,
			bar width=1cm,
			ylabel={Frequency},
                xlabel={Common Weakness Enumeration},
			symbolic x coords={other,119,120,476,122,190,121,78,787,20,762},
			x tick label style={rotate=45, anchor=north, align=left},
			nodes near coords,
			nodes near coords align={vertical}          
			]
   			\addplot[blue!85,fill] coordinates {(other,66761)};
			\addplot[blue!75,fill] coordinates {(119,25011)};
			\addplot[blue!65,fill] coordinates {(120,22213)};
			\addplot[blue!60,fill] coordinates {(476,11499)};
			\addplot[blue!55,fill] coordinates {(122,3509)};
			\addplot[blue!50,fill] coordinates {(190,3307)};
			\addplot[blue!45,fill] coordinates {(121,3041)};
			\addplot[blue!40,fill] coordinates {(78,2872)};
            \addplot[blue!35,fill] coordinates {(787,2621)};
			\addplot[blue!30,fill] coordinates {(20,2200)};
			\addplot[blue!25,fill] coordinates {(762,2072)};    

        \legend{CWE-119: Improper Restriction ,
                CWE-120: Buffer Copy,
                CWE-476: NULL Pointer Dereference ,
                CWE-122: Heap-based Buffer Overflow,
                CWE-190 :  Integer Overflow or Wraparound,
                CWE-121 : Stack-based Buffer Overflow ,
                CWE-78 : OS Command Injection,
                CWE-787 : Out-of-bounds Write ,
                CWE-20 : Improper Input Validation ,
                CWE-762 : Mismatched Memory Management}
		\end{axis}

		\end{tikzpicture}   
}
        \caption{\color{black}Top 11 most frequent vulnerability categories in FalconVulnDB.}
        \label{fig:dist_graph_Agg}
	\end{figure}
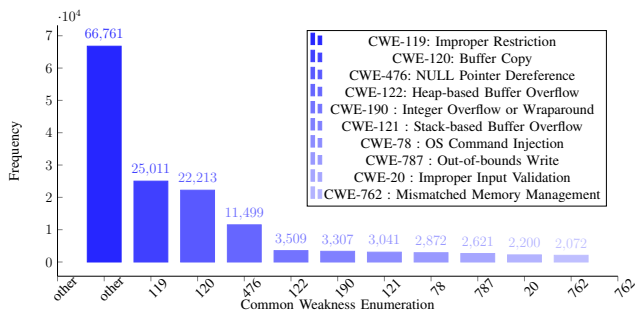
  
\end{center}

\subsubsection{\textcolor{black}{Data Integration, Standardization, and Bias Mitigation}}

\textcolor{black}{We acknowledge the known data quality issues associated with widely used datasets like BigVul, as highlighted in prior studies such as LineVul \cite{fu2022linevul}. To ensure the reliability of our results, we implemented a rigorous pre-processing pipeline that included deduplication to eliminate redundant records, validation of vulnerability labels by cross-referencing with publicly available vulnerability descriptions, and removing incomplete or ambiguous entries.}

\textcolor{black}{We employed a structured integration and standardization process to ensure consistency, reliability, and transparency in combining these heterogeneous sources. First, we established a unified schema that standardized fields such as vulnerability labels, CWE tags, and severity metrics. For datasets that provided direct CWE mappings, we incorporated them as-is; for those offering only CVE references or binary vulnerability labels (e.g., ReVeal), we cross-checked these identifiers against authoritative databases (e.g., NVD) to map or infer the most likely CWE categories whenever possible. For instance, if a function marked as vulnerable in ReVeal referenced a CVE later confirmed by NVD as related to a buffer overflow vulnerability, we annotated that sample with the corresponding CWE (e.g., CWE-120). When no reliable CWE classification could be assigned, we preserved the binary vulnerability label to maintain dataset diversity.}

\textcolor{black}{We also addressed overlaps and conflicts among datasets. In one example, the same function appeared in both Big-Vul and SySeVR, but SySeVR indicated it had been patched, while Big-Vul still listed it as vulnerable. We adopted a hierarchical decision approach in such cases: we consulted NVD and official vendor advisories to verify the patch status. If confirmed, the function was recorded as patched and no longer considered an active vulnerability. Another example involved duplicates in which multiple datasets reported the same CVE but provided slightly different severity scores. Here, we normalized the severity by referencing a standard scale (e.g., CVSS) and took the median of reported severity metrics, ensuring a more robust and standardized representation of the vulnerability’s impact.}

\textcolor{black}{To reduce bias, we examined the data distribution across various dimensions—such as vulnerability type, programming language, and software domain—and identified areas of overrepresentation. For example, if stack-based buffer overflows (CWE-121) were disproportionately frequent due to their abundance in a particular dataset, we incorporated more diverse examples from other sources or filtered out duplicates to achieve a more balanced distribution. Additionally, we validated portions of FalconVulnDB against established benchmarks like NVD. For example, if our integrated dataset suggested null pointer dereferences (CWE-476) were exceedingly rare compared to NVD distributions, we revisited our filtering and inclusion criteria to better align with real-world data.}

\section{Performance Evaluation} \label{sec:evaluation}

\begin{figure*}[htbp] 
\centering
\includegraphics[width=0.9\textwidth]{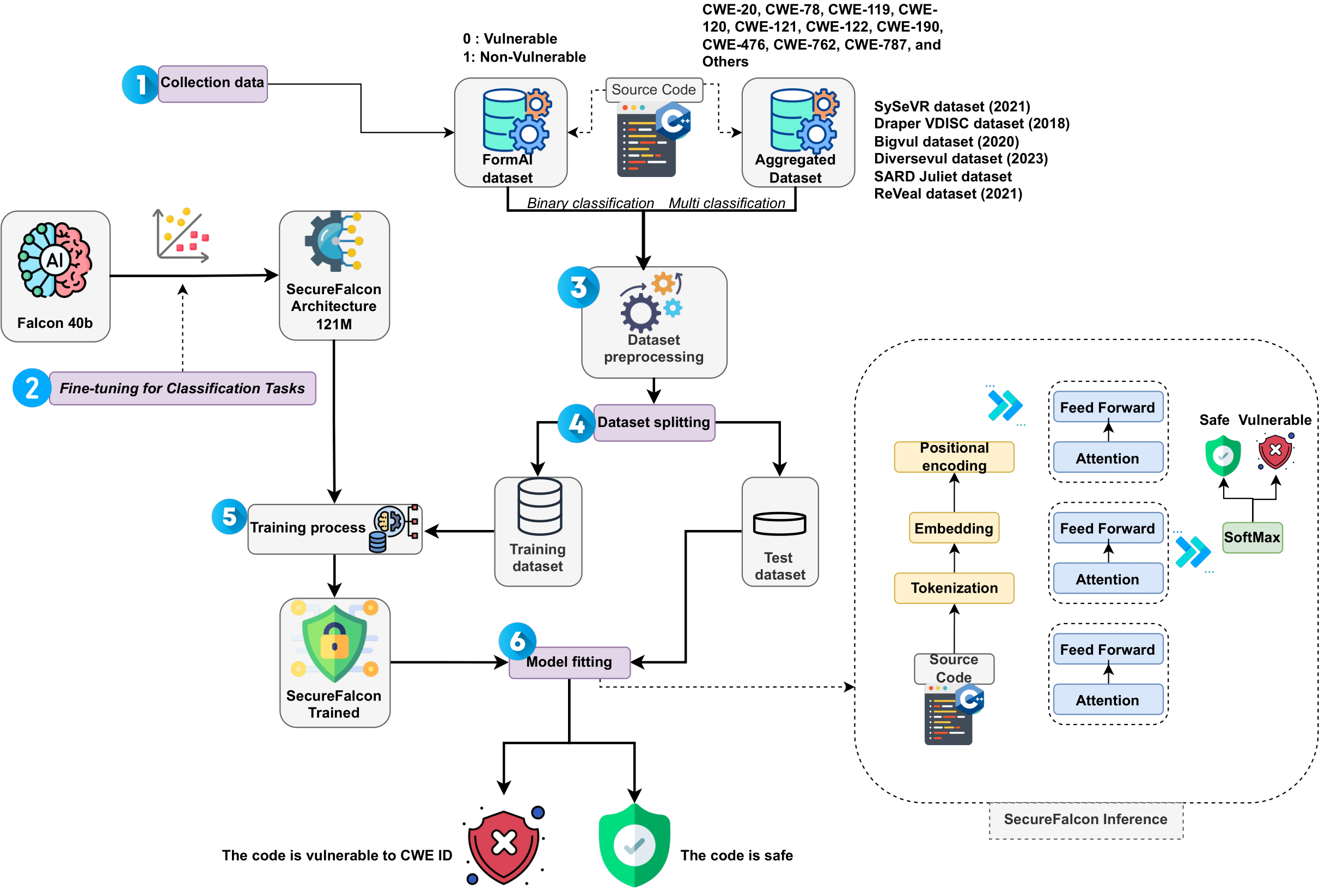}
\caption{Performance evaluation steps of \textit{SecureFalcon} model.}
\label{fig:fig2}
\end{figure*}
Performance evaluation of \texttt{SecureFalcon} model for software vulnerability detection involves a series of steps, as presented in Fig. \ref{fig:fig2}. First, we collect a large and diverse dataset of software code from the FormAI and the aggregated datasets (FalconVulnDB), where instances of vulnerable and non-vulnerable code are correctly labeled. Once we have the datasets, we pre-process them by transforming the raw source code into a format suitable for the \texttt{SecureFalcon} model, tokenizing the code into appropriate separate sub-strings. Next, we initiate the \texttt{SecureFalcon} model with the desired architecture using the Transformers library. Then, we define the training and evaluation settings, which involve setting various hyperparameters like learning rate, batch size, and number of epochs. We start the fine-tuning process, where the model will learn from the labeled dataset to differentiate between vulnerable and non-vulnerable code. We evaluate the model regularly during training to track its progress and adjust hyperparameters when necessary. After training the model, we test it on separate data amalgamated from the different datasets to verify its effectiveness. Lastly, after fine-tuning, the resulting model should be able to predict whether a given piece of code is vulnerable or not. 

\subsection{Experimental Setup}
TABLE~\ref{tab:tab3} outlines the experimental setup and parameters used for fine-tuning \texttt{SecureFalcon} for software security.

\begin{table}[ht!]
\centering
\begin{tcolorbox}[colback=gray!10]
\caption{Configuration of \texttt{SecureFalcon}.}
\label{tab:tab3}
\begin{tabular}{ll}
\hline
\textbf{Parameter} & \textbf{Value} \\\hline
Pretrained model ID & FalconLLM 40B  \\\hline
Number of parameters  &  121M \\\hline
Hidden size & 768 \\\hline
Number of hidden layers & 12 \\\hline
Number of attention heads & 12 \\\hline
Intermediate-size & 3072 \\\hline
Maximum position embeddings & 514 \\\hline
Number of labels & 2/12 \\\hline
Tokenizer padding side & Left \\\hline
Padding token ID & 11 \\\hline
BOS token ID & 11 \\\hline
EOS token ID & 11 \\\hline
Maximum length of tokens & 2048 \\\hline
Tokenizer truncation & True \\\hline
Tokenizer padding & True \\\hline
Return tensor format & PyTorch tensors ('pt') \\\hline
Return token type IDs & False \\\hline
Return attention mask & True \\\hline
Batch size & 256 \\\hline
GPU & 32 A100 40GB\\\hline
Optimizer & AdamW \\\hline
AdamW Learning Rate (LR) &  2e-2 and 2e-5 \\\hline
AdamW Epsilon & 1e-8 \\\hline
Number of training epochs & 10 \\\hline
Early stopping & Enabled (patience=3) \\\hline
Random seed value & 42 \\\hline
Maximum gradient norm & 1.0 \\\hline
Loss computation & Cross-entropy loss \\\hline
Hidden activation function & "gelu" \\\hline
Initializer range & 0.02 \\\hline
Layer norm epsilon & 1e-5 \\\hline
Attention probs dropout prob & 0.1 \\\hline
Hidden dropout prob & 0.1 \\\hline
Torch\_dtype & float32 \\\hline
Transformers\_version  & 4.30.2\\\hline
SMDDP\_version & 2.1.0 \\\hline
Pytorch\_version & 2.1.0 \\\hline
Python\_version & 3.10 \\\hline
\end{tabular}
\end{tcolorbox}
\end{table}

The \texttt{SecureFalcon} model features 121 million parameters, a hidden size of 768, twelve hidden layers, and twelve attention heads. This model employs an intermediate size of 3072 and supports up to 514 position embeddings. Tokenization was performed using a left-padding approach, with a padding token ID, BOS (beginning of sentence) token ID, and EOS (end of sentence) token ID all set to 11. The tokenizer settings enforced truncation and padding, producing PyTorch tensor outputs and including an attention mask, while token type IDs were not returned. We set the maximum token length to 2048. For training, batch sizes were set to 256, utilizing 32 Nvidia A100 40GB GPUs to ensure efficient processing. We employed the AdamW optimizer with learning rates set at 2e-2 and 2e-5 and an epsilon value of 1e-8. The training was carried out over ten epochs to prevent overfitting, with early stopping enabled after three epochs without improvement. For loss computation, we utilized cross-entropy loss. The model was evaluated based on accuracy, precision, recall, and the F1 score, with metrics averaged using the 'micro' method to reflect the contribution of each instance to the overall metric. The additional configuration included setting the hidden activation function to GELU, an initializer range of 0.02, and a layer norm epsilon of 1e-5. The dropout probabilities for attention and hidden layers were set to $0.1$. The models were run under a float32 torch data type setting, ensuring compatibility and optimal performance on the specified transformer library version 4.30.2.

\subsection{Experimental Results}

Tables~\ref{tab:classification_report1} and~\ref{tab:classification_report2} present the classification report of \texttt{SecureFalcon}-121M with LR = 2e-5 and LR = 2e-2. With an LR of 2e-5, the model yields the highest accuracy of $0.94$, supported by a high precision, recall, and F1-score for both classes (0.89, 0.84, 0.86 for `NOT VULNERABLE' and 0.95, 0.97, 0.96 for `VULNERABLE') (see Tables~\ref{tab:classification_report1})

\begin{table}[ht]
\centering
\caption{Classification report of \texttt{SecureFalcon} 121M with LR = 2e-5 using FormAI dataset.}
\label{tab:classification_report1}
\begin{tabular}{lcccc}
\hline
 & Precision & Recall & F1-Score & Support\\
\hline
0 & 0.89 & 0.84 & 0.86 & 4528\\ 
1 & 0.95 & 0.97 & 0.96 & 15533\\\hline

Accuracy & \multicolumn{4}{c}{0.94}\\ \hline
Macro avg & 0.92 & 0.90 & 0.91 & 20061\\
Weighted avg & 0.94 & 0.94 & 0.94 & 20061\\
\hline
\end{tabular}\\
\vspace{0.3em}
0: NOT VULNERABLE, 1: VULNERABLE 
\end{table}

\begin{table}[ht]
\centering
\caption{Classification report of \texttt{SecureFalcon} 121M with LR = 2e-2 using FormAI dataset.}
\label{tab:classification_report2}
\begin{tabular}{lcccc}
\hline
 & Precision & Recall & F1-Score & Support\\
\hline
0 & 0.67 & 0.80 & 0.73 & 4528\\ 
1 & 0.94 & 0.88 & 0.91 & 15533\\\hline

Accuracy & \multicolumn{4}{c}{0.87} \\ \hline
Macro avg & 0.80 & 0.84 & 0.82 & 20061\\
Weighted avg & 0.88 & 0.87 & 0.87 & 20061\\
\hline
\end{tabular}\\
\vspace{0.3em}
0: NOT VULNERABLE, 1: VULNERABLE 
\end{table}This performance considerably drops when the learning rate is increased to 2e-2. Although the precision for `VULNERABLE' remains high ($0.94$), the `NOT VULNERABLE' metrics take a substantial hit, decreasing overall accuracy to $0.87$. 

Tables~\ref{tab:simple_metrics1},~\ref{tab:simple_metrics2},~\ref{tab:simple_metrics4} present the training and validation accuracy and loss over several epochs for \texttt{SecureFalcon} model with differing configurations. With a Learning Rate (LR) of 2e-5 (Table~\ref{tab:simple_metrics1}), accuracy consistently increases, and loss decreases over the epochs for both training and validation. However, with an LR of 2e-2 (Table~\ref{tab:simple_metrics2}), lower accuracy and higher loss indicate that a larger learning rate could lead to sub-optimal learning.
\begin{table}[ht]
\centering
\caption{Training and Validation Accuracy and Loss Across Epochs of \texttt{SecureFalcon} 121M with LR = 2e-5 using FormAI dataset.}
\label{tab:simple_metrics1}
\begin{tabular}{lcccc}
\hline
& \multicolumn{2}{c}{Training} & \multicolumn{2}{c}{Validation} \\
\cline{2-3} \cline{4-5}
Epoch & Accuracy & Loss & Accuracy & Loss \\
\hline
1 & 0.82 & 0.39 & 0.88 & 0.27 \\
2 & 0.87 & 0.29 & 0.89 & 0.25 \\
3 & 0.89 & 0.25 & 0.90 & 0.23 \\
4 & 0.91 & 0.21 & 0.91 & 0.21 \\
5 & 0.93 & 0.17 & 0.93 & 0.19 \\
6 & 0.95 & 0.12 & 0.93 & 0.19 \\
7 & 0.97 & 0.09 & 0.94 & 0.19 \\
\hline
\end{tabular}
\end{table}

\begin{table}[ht]
\centering
\caption{Training and Validation Accuracy and Loss Across Epochs of \texttt{SecureFalcon} 121M with LR= 2e-2 using FormAI dataset.}
\label{tab:simple_metrics2}
\begin{tabular}{lcccc}
\hline
& \multicolumn{2}{c}{Training} & \multicolumn{2}{c}{Validation} \\
\cline{2-3} \cline{4-5}
Epoch & Accuracy & Loss & Accuracy & Loss \\
\hline
1 & 0.70 & 0.61 & 0.82 & 0.43 \\
2 & 0.78 & 0.48 & 0.80 & 0.42 \\
3 & 0.80 & 0.44 & 0.85 & 0.34 \\
4 & 0.83 & 0.38 & 0.81 & 0.41 \\
5 & 0.82 & 0.40 & 0.86 & 0.31 \\
6 & 0.84 & 0.36 & 0.86 & 0.30 \\
7 & 0.85 & 0.34 & 0.86 & 0.30 \\
8 & 0.85 & 0.34 & 0.87 & 0.30 \\
\hline
\end{tabular}
\end{table}

\begin{table}[ht]
\centering
\caption{Training and Validation Accuracy and Loss Across Epochs of \texttt{SecureFalcon} 121M with LR = 2e-5 using FalconVulnDB dataset}
\label{tab:simple_metrics4}
\begin{tabular}{lcccc}
\hline
& \multicolumn{2}{c}{Training} & \multicolumn{2}{c}{Validation} \\
\cline{2-3} \cline{4-5}
Epoch & Accuracy & Loss & Accuracy & Loss \\
\hline
1 & 0.91 & 0.30 & 0.91 & 0.31 \\
2 & 0.92 & 0.26 & 0.91 & 0.32 \\
\hline
\end{tabular}
\end{table}

\begin{figure}[ht]
  \centering
  \begin{subfigure}[b]{0.35\textwidth}
    \includegraphics[width=\textwidth]{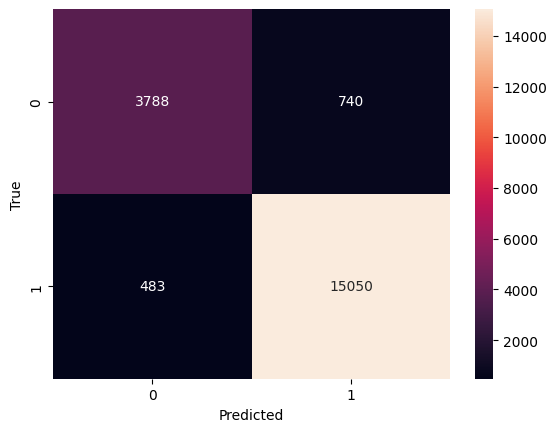}
    \caption{\texttt{SecureFalcon} with LR = 2e-5}
    \label{fig:imageb}
  \end{subfigure}
  \begin{subfigure}[b]{0.35\textwidth}
    \includegraphics[width=\textwidth]{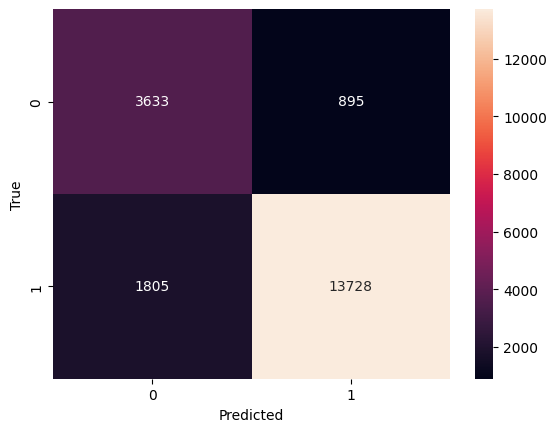}
    \caption{\texttt{SecureFalcon} with LR = 2e-2}
    \label{fig:imagec}
  \end{subfigure}\\
  \vspace{0.3em}
  0: NOT VULNERABLE, 1: VULNERABLE 

    \caption{Confusion matrix of \texttt{SecureFalcon} 121M classification using FormAI dataset.}
  \label{fig:fig3}
  \end{figure}

Figure~\ref{fig:fig3} illustrates the confusion matrix for the binary classification performed by \texttt{SecureFalcon}. From the provided confusion matrices, we can make some observations regarding the performance of the \texttt{SecureFalcon} models with different configurations. A decrease in performance is observed when the learning rate is raised to 2e-2. The number of false negatives drastically increased from $483$ to $1805$, with false positives rising from $740$ to $895$. Correspondingly, the true positives and negatives decreased from $15050$ to $13728$ and from $3788$ to $3633$, respectively. This shows that a larger learning rate, in this case, led to more misclassifications, indicating an over-optimistic learning process that possibly led to overfitting or instability during training.  This is due to the high learning rate contributing to more drastic changes in weight, leading to instability or overfitting.

The multi-classification report of the \texttt{SecureFalcon} model, as presented in Table~\ref{tab:classification_report4}, exhibits a comprehensive evaluation of the model's performance across various Common Weakness Enumerations (CWEs) on FalconVulnDB. 
We conducted several experiments to determine the model's rationale for the classification. One of the interpretability tools\footnote{\url{https://github.com/cdpierse/transformers-interpret}  (accessed 9 May 2024)} used was a library highlighting the distribution of attention, indicating some discrepancies in our model results. To enhance the knowledge of the model of the syntactic and lexical nature of the programming language, we embed the tokens shown in Table~\ref{tab:spec_token}. The tokens are keywords, punctuation, and API calls defined in the C/C++ manual 
and its extended libraries\footnote{\url{https://www.gnu.org/software/libc/manual/pdf/libc.pdf}    (accessed 15 May 2024)}\footnote{\url{https://learn.microsoft.com/en-us/cpp/cpp/cpp-language-reference}   (accessed 15 May 2024)}, 

an approach similar to work \cite{hanif2022vulberta}. Testing with the new tokens included in the fine-tuning, in turn, updating the embedding layer has yielded higher precision and recall. 
This is due to maintaining the syntax of the pre-defined tokens rather than going through the subword tokenization process. As such, the meaning of these essential pre-defined tokens in the source code is preserved.

\begin{table}[ht]
\centering
\centering
\caption{\color{black}Classification report of \textit{SecureFalcon} 121M with LR = 2e-5 using FalconVulnDB dataset.}
\label{tab:classification_report4}
\begin{tabular}{lcccc}
\hline
& Precision & Recall & F1-Score & Support \\
\hline
Not-Vulnerable & 0.93 & 0.97 & 0.95 & 124355 \\
CWE-20 & 0.46 & 0.14 & 0.22 & 440 \\
CWE-78 & 1.00 & 0.98 & 0.99 & 574 \\
CWE-119 & 0.75 & 0.75 & 0.75 & 5002 \\
CWE-120 & 0.63 & 0.84 & 0.72 & 4443 \\
CWE-121 & 0.99 & 0.99 & 0.99 & 608 \\
CWE-122 & 1.00 & 0.99 & 0.99 & 702 \\
CWE-190 & 0.97 & 0.78 & 0.87 & 662 \\
CWE-476 & 0.64 & 0.45 & 0.53 & 2300 \\
CWE-762 & 1.00 & 1.00 & 1.00 & 414 \\
CWE-787 & 0.27 & 0.26 & 0.26 & 524 \\
Other & 0.89 & 0.54 & 0.67 & 13352 \\
\hline
\multicolumn{5}{c}{ } \\
\hline
Accuracy & \multicolumn{4}{c}{0.92} \\
\hline
Macro Avg & 0.79 & 0.72 & 0.74 & 153376 \\
Weighted Avg & 0.91 & 0.91 & 0.90 & 153376 \\
\hline
\end{tabular}\\
\vspace{0.3em}
\end{table}

The model demonstrates high precision and recall for most categories, particularly excelling in identifying non-vulnerable instances with a precision of $0.93$, recall of $0.97$, and an F1-score of $0.95$, indicating robustness in distinguishing non-vulnerable cases. Remarkably, it achieves perfect or near-perfect performance in identifying CWE-78, CWE-121, CWE-122, and CWE-762 vulnerabilities, showcasing its effectiveness in detecting specific types of vulnerabilities with high accuracy. However, the model shows limitations in recognizing certain vulnerabilities, notably CWE-20 and CWE-787, with notably lower precision and recall values, suggesting areas for improvement in future iterations. The low F1-score for CWE-20 ($0.22$) and CWE-787 ($0.26$) highlights the model's challenge in accurately classifying these vulnerabilities, possibly due to the complexity of the patterns associated with these CWEs or a limited representation in the training data.

\begin{figure}[htbp] 
\centering
\includegraphics[width=0.5\textwidth]{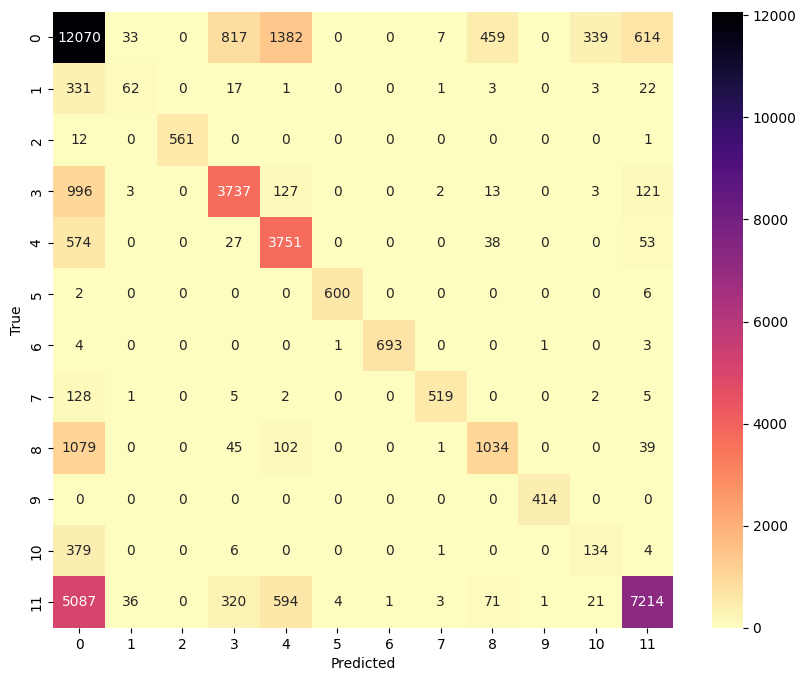}
\caption{\color{black}Confusion Matrix of \texttt{SecureFalcon} 121M Multi-classification using FalconVulnDB Dataset.}
\label{fig:confisuion_matrix}
\end{figure}

The aggregated accuracy of $0.92$ and the weighted average precision, recall, and F1 score reflect the model's high competence in vulnerability classification across diverse vulnerabilities. However, the varying performance across different CWEs underscores the importance of continued model refinement and targeted training to enhance the model's sensitivity and specificity, particularly for those vulnerabilities where it currently underperforms. \textcolor{black}{The confusion matrix for multiclass classification on the FalconVulnDB Dataset is shown in Figure~\ref{fig:confisuion_matrix}.}

\textcolor{black}{As evident in the matrix, the model demonstrates strong performance in detecting the "Not-Vulnerable" class, with a high number of true positives and minimal misclassifications. However, for certain challenging vulnerability types, such as CWE-20 (Improper Input Validation) and CWE-787 (Out-of-Bounds Write), the model struggles to achieve the same level of precision and recall. This is likely due to the limited support for these classes (440 and 524 samples, respectively) in the dataset, leading to a higher rate of misclassification into more frequent categories, such as CWE-119 (Memory Corruption) or the "Other" class.}

\textcolor{black}{Furthermore, rare and complex classes like CWE-476 (NULL Pointer Dereference) and CWE-190 (Integer Overflow or Wraparound) show lower performance compared to well-represented classes, such as CWE-78 (OS Command Injection) and CWE-121 (Stack-Based Buffer Overflow), which have higher support and more distinct patterns. The confusion matrix highlights these discrepancies, offering valuable insights into the model's misclassification patterns. For future work, we plan to enhance the dataset by augmenting samples for underrepresented classes and leveraging advanced techniques such as the focal loss \cite{mukhoti2020calibrating} to prioritize learning on these challenging categories. These improvements will help address the current limitations and improve the model's robustness for detecting rare vulnerabilities.}

\begin{table}
\centering
\caption{List of special tokens added in the Falcon40b tokenizer}
\label{tab:spec_token}
\begin{tblr}{
   column{2} = {c},
   cell{1}{3} = {c},
   cell{2}{3} = {c},
   cell{3}{3} = {c},
   hline{2,5} = {1-3}{},
 }
 Tokens      & Count & Examples              &  \\
 Punctuation & 72    & !=, ++, =             &  \\
 Keywords    & 123   & char, const, continue &  \\
 API calls   & 394   & malloc, strncpy, atoi &  
 \end{tblr}
 \end{table}

\textcolor{black}{
\begin{table}[ht]
\centering
\caption{\textcolor{black}{Comprehensive Evaluation Metrics for \textit{SecureFalcon} 121M with LR = 2e-5 using the FalconVulnDB Dataset.}}
\label{tab:comprehensive_metrics}
\begin{tabular}{lc}
\toprule
\textbf{\textcolor{black}{Metric}} & \textbf{\textcolor{black}{Value}} \\
\midrule
\textcolor{black}{Cohen’s Kappa} & \textcolor{black}{0.86} \\
\textcolor{black}{MCC (Matthews Corr. Coeff.)} & \textcolor{black}{0.79} \\
\textcolor{black}{Macro F1-Score} & \textcolor{black}{0.74} \\
\textcolor{black}{Weighted F1-Score} & \textcolor{black}{0.90} \\
\textcolor{black}{ROC-AUC (Macro Avg)} & \textcolor{black}{0.88} \\
\textcolor{black}{PR-AUC (Macro Avg)} & \textcolor{black}{0.85} \\
\textcolor{black}{Accuracy} & \textcolor{black}{0.92} \\
\textcolor{black}{Macro Precision} & \textcolor{black}{0.79} \\
\textcolor{black}{Macro Recall} & \textcolor{black}{0.72} \\
\textcolor{black}{Weighted Precision} & \textcolor{black}{0.91} \\
\textcolor{black}{Weighted Recall} & \textcolor{black}{0.91} \\
\textcolor{black}{Specificity (Macro Avg)} & \textcolor{black}{0.89} \\
\textcolor{black}{Log Loss} & \textcolor{black}{0.35} \\
\textcolor{black}{Brier Score} & \textcolor{black}{0.15} \\
\textcolor{black}{Hamming Loss} & \textcolor{black}{0.08} \\
\bottomrule
\end{tabular}
\end{table}
}

\subsection{\textcolor{black}{Comprehensive Evaluation Metrics}}

\textcolor{black}{Table~\ref{tab:comprehensive_metrics} presents a comprehensive evaluation of the \texttt{SecureFalcon} 121M model trained with a learning rate of \(2 \times 10^{-5}\) on the FalconVulnDB dataset, showcasing robust performance across multiple metrics. The following metrics are used in our analysis:}

\begin{itemize}
    \item \textcolor{black}{
        Cohen’s Kappa: Measures the agreement between the predicted and actual classifications, accounting for the possibility of agreement occurring by chance.
        \begin{equation}
            \kappa = \frac{P_o - P_e}{1 - P_e}
        \end{equation}
        where \( P_o \) is the observed agreement and \( P_e \) is the expected agreement by chance.
    }
    
    \item \textcolor{black}{
       MCC (Matthews Correlation Coefficient): Evaluates the quality of binary classifications, considering true and false positives and negatives.
        \begin{equation}
            \text{MCC} = \frac{TP \times TN - FP \times FN}{\sqrt{(TP + FP)(TP + FN)(TN + FP)(TN + FN)}}
        \end{equation}
     where $TP$ (True Positive) and $TN$ (True Negative) represent the correctly predicted positive and negative instances, respectively, while $FP$ (False Positive) and $FN$ (False Negative) denote the incorrectly predicted positive and negative instances.
    }
    
    \item \textcolor{black}{
        Macro F1-Score: The unweighted average of F1-scores for each class, treating all classes equally.
        \begin{equation}
            \text{F1}_{\text{macro}} = \frac{1}{C} \sum_{i=1}^{C} \frac{2 \times P_i \times R_i}{P_i + R_i}
        \end{equation}
        where \( C \) is the number of classes, \( P_i \) is the precision, and \( R_i \) is the recall for class \( i \).
    }
    
    \item \textcolor{black}{
        Weighted F1-Score: The average of F1-scores weighted by the number of true instances for each class.
        \begin{equation}
            \text{F1}_{\text{weighted}} = \sum_{i=1}^{C} w_i \times \text{F1}_i
        \end{equation}
        where \( w_i = \frac{\text{Support}_i}{\sum_{j=1}^{C} \text{Support}_j} \).
    }
    
    \item \textcolor{black}{
        ROC-AUC (Macro Avg): The average Area Under the Receiver Operating Characteristic Curve across all classes, assessing the model's ability to distinguish between classes.
        \begin{equation}
            \text{ROC-AUC}_{\text{macro}} = \frac{1}{C} \sum_{i=1}^{C} \text{ROC-AUC}_i
        \end{equation}
    }
    
    \item \textcolor{black}{
        PR-AUC (Macro Avg): The average Area Under the Precision-Recall Curve across all classes, focusing on the trade-off between precision and recall.
        \begin{equation}
            \text{PR-AUC}_{\text{macro}} = \frac{1}{C} \sum_{i=1}^{C} \text{PR-AUC}_i
        \end{equation}
    }
    
    \item \textcolor{black}{
        Macro Precision: The unweighted average precision across all classes, emphasizing the model's ability to correctly identify positive instances.
        \begin{equation}
            \text{Precision}_{\text{macro}} = \frac{1}{C} \sum_{i=1}^{C} P_i
        \end{equation}
    }
    
    \item \textcolor{black}{
        Macro Recall: The unweighted average recall across all classes, highlighting the model's ability to capture all relevant instances.
        \begin{equation}
            \text{Recall}_{\text{macro}} = \frac{1}{C} \sum_{i=1}^{C} R_i
        \end{equation}
    }
    
    \item \textcolor{black}{
        Weighted Precision: The precision score weighted by the number of true instances for each class, reflecting overall precision performance.
        \begin{equation}
            \text{Precision}_{\text{weighted}} = \sum_{i=1}^{C} w_i \times P_i
        \end{equation}
    }
    
    \item \textcolor{black}{
        Weighted Recall: The recall score weighted by the number of true instances for each class, indicating overall recall performance.
        \begin{equation}
            \text{Recall}_{\text{weighted}} = \sum_{i=1}^{C} w_i \times R_i
        \end{equation}
    }
    
    \item \textcolor{black}{
       Specificity (Macro Avg): The average specificity across all classes, measuring the model's ability to correctly identify negative instances.
        \begin{equation}
            \text{Specificity}_{\text{macro}} = \frac{1}{C} \sum_{i=1}^{C} \frac{TN_i}{TN_i + FP_i}
        \end{equation}
    }
    
    \item \textcolor{black}{
        Log Loss: Evaluates the uncertainty of the model's predictions based on the probability estimates.
        \begin{equation}
            \text{Log Loss} = -\frac{1}{N} \sum_{i=1}^{N} \left[ y_i \log(p_i) + (1 - y_i) \log(1 - p_i) \right]
        \end{equation}
        where \( N \) is the number of samples, \( y_i \) is the true label, and \( p_i \) is the predicted probability.
    }
    
    \item \textcolor{black}{
        Brier Score: Measures the mean squared difference between predicted probabilities and actual outcomes.
        \begin{equation}
            \text{Brier Score} = \frac{1}{N} \sum_{i=1}^{N} (p_i - y_i)^2
        \end{equation}
    }
    
    \item \textcolor{black}{
        Hamming Loss: The fraction of incorrect labels to the total number of labels.
        \begin{equation}
            \text{Hamming Loss} = \frac{1}{N} \sum_{i=1}^{N} \mathbb{1}(y_i \neq \hat{y}_i)
        \end{equation}
    }
\end{itemize}

\textcolor{black}{A high Accuracy of 0.92 and a Weighted F1-Score of 0.90 suggest strong overall effectiveness and good handling of class imbalances. Cohen’s Kappa of 0.86 and MCC of 0.79 underscore the reliability and stability of the model’s predictions. At the same time, the ROC-AUC (Macro Avg) of 0.88 and PR-AUC (Macro Avg) of 0.85 demonstrate its discriminative power and capacity to maintain solid precision and recall across classes.}

\textcolor{black}{The differences between the macro- and weighted-average metrics (e.g., Macro F1-Score of 0.74 versus Weighted F1-Score of 0.90) highlight that certain less frequent classes may not be identified as accurately as more common ones. The Log Loss of 0.35 and Brier Score of 0.15 indicate relatively good calibration, suggesting that the model’s probability estimates are reasonably aligned with observed outcomes. Additionally, a Hamming Loss of 0.08 and a Specificity (Macro Avg) of 0.89 confirm that the model makes relatively few incorrect predictions and can identify negative instances correctly. These findings suggest that, while \textit{SecureFalcon} performs admirably, further improvements—such as targeted data augmentation or calibration adjustments—could enhance its robustness across all vulnerability categories.
}

\subsection{Comparison of \texttt{SecureFalcon} with LLM, ML, and DL models}

Table~\ref{tab:classification_report_extended} compares various machine learning (ML) models and their accuracy in performing multiclass and binary classification tasks on the FormAI and FalconVulnDB datasets. Large Language Models (LLMs) such as RoBERTa, BERT, CodeBERT, and \texttt{SecureFalcon} show a notable performance difference compared to traditional ML models for multiclass classification. \texttt{SecureFalcon}, in particular, stands out with the highest accuracy of 0.92 on both datasets, significantly outperforming other LLM models like RoBERTa and BERT, which achieve accuracies of around 0.70-0.85. Among the traditional ML models, Random Forest (RF) performs the best with accuracies of 0.77 on the FormAI dataset and 0.81 on the FalconVulnDB dataset. However, it still lags behind the LLM models.

The trend remains similar in binary classification, with \texttt{SecureFalcon} leading with an accuracy of 0.94 on the FormAI dataset and 0.92 on the FalconVulnDB dataset. Other LLM models, such as RoBERTa, BERT, and CodeBERT, also perform well, achieving accuracies between 0.81 and 0.89 across both datasets. Traditional ML models show competitive results in this task, with Random Forest achieving an accuracy of 0.91 on the FormAI dataset and 0.89 on the FalconVulnDB dataset, which is relatively close to the performance of some LLMs. However, the consistently superior performance of \texttt{SecureFalcon} highlights the advantages of fine-tuning large models specifically for software vulnerability detection tasks, making it a robust choice for both multiclass and binary classification in cybersecurity contexts.

\textcolor{black}{Can traditional machine learning models effectively capture vulnerability patterns without sophisticated semantic understanding? For the FormAI dataset, which is synthetic and constructed in a manner favoring simpler patterns, the answer appears to be yes. In the multiclass scenario on the FormAI dataset, CodeBERT—a general pre-trained LLM fine-tuned for classification—achieves an accuracy of 0.71, while specific traditional ML models, such as Random Forest (RF), reach up to 0.77. This suggests that vulnerability patterns in FormAI can be successfully modeled using classical feature engineering or straightforward pattern recognition. Traditional ML methods, often relying on syntactic features (e.g., token frequencies or specific keywords), perform competitively because well-established, surface-level indicators may represent the dataset’s vulnerabilities.}

\textcolor{black}{But what happens when the complexity of vulnerabilities increases, demanding deeper semantic comprehension? In the FalconVulnDB dataset, where vulnerabilities involve more subtle and context-dependent cues, traditional ML models show a performance plateau (e.g., RF at 0.81 accuracy in multiclass classification). In contrast, LLM-based models—pre-trained on extensive corpora and adept at contextualizing code semantics—excel (e.g., CodeBERT at 0.88). The heightened complexity and nuance of FalconVulnDB vulnerabilities seemingly require the richer representational capacity and more advanced reasoning capabilities that LLMs provide.}

\textcolor{black}{Table \ref{tab:comparisondeep} compares our proposed \texttt{SecureFalcon} model against several state-of-the-art deep learning-based vulnerability detection methods, including SySeVR~\cite{li2021sysevr}, VulDeePecker~\cite{Li2018vuldeepecker}, and Devign~\cite{zhou2019devign}. While the earlier approaches often rely on hand-crafted features (e.g., syntax characteristics or code gadgets) and are typically limited to detecting a narrower set of vulnerabilities, \texttt{SecureFalcon} leverages a large language model architecture derived from Falcon-40B. This allows \texttt{SecureFalcon} to incorporate both synthetically generated data and real-world vulnerability instances drawn from a broad set of sources (FalconVulnDB), ultimately enabling it to target a far more comprehensive range of CWE types, including the industry-recognized Top 25 Most Dangerous CWEs \footnote{\url{https://cwe.mitre.org/top25/archive/2024/2024_cwe_top25.html}}. In contrast, the earlier methods examined focus on a smaller subset of vulnerabilities and do not integrate synthetic data to enhance coverage and robustness.}

\begin{table*}[!ht]
\centering
{\color{black}
\caption{\textcolor{black}{Comparison of \texttt{SecureFalcon} with LLM models and traditional machine learning using the FormAI and FalconVulnDB datasets.}}
\label{tab:classification_report_extended}

\begin{tabular}{|c|c|c|cccc|cccc|}
\hline
\multirow{3}{*}{\textbf{\textcolor{black}{Task}}} & \multirow{3}{*}{\textbf{\textcolor{black}{ML type}}} & \multirow{3}{*}{\textbf{\textcolor{black}{Model}}} & \multicolumn{4}{c|}{\textbf{\textcolor{black}{FormAI dataset}}}                                                                                         & \multicolumn{4}{c|}{\textbf{\textcolor{black}{FalconVulnDB dataset}}}                                                                                     \\ \cline{4-11} 
 &  &  & \multicolumn{1}{c|}{\textbf{\textcolor{black}{Acc}}} & \multicolumn{1}{c|}{\textbf{\textcolor{black}{Prec}}} & \multicolumn{1}{c|}{\textbf{\textcolor{black}{Rec}}} & \textbf{\textcolor{black}{F1}} & \multicolumn{1}{c|}{\textbf{\textcolor{black}{Acc}}} & \multicolumn{1}{c|}{\textbf{\textcolor{black}{Prec}}} & \multicolumn{1}{c|}{\textbf{\textcolor{black}{Rec}}} & \textbf{\textcolor{black}{F1}} \\ \hline

\multirow{11}{*}{\textcolor{black}{Multiclass}} 
& \multirow{4}{*}{\textcolor{black}{LLM models}} 
& \textcolor{black}{RoBERTa} 
& \multicolumn{1}{c|}{\textcolor{black}{0.70}} & \multicolumn{1}{c|}{\textcolor{black}{0.72}} & \multicolumn{1}{c|}{\textcolor{black}{0.68}} & \textcolor{black}{0.70} 
& \multicolumn{1}{c|}{\textcolor{black}{0.85}} & \multicolumn{1}{c|}{\textcolor{black}{0.87}} & \multicolumn{1}{c|}{\textcolor{black}{0.84}} & \textcolor{black}{0.85} \\ \cline{3-11} 

&  & \textcolor{black}{BERT} 
& \multicolumn{1}{c|}{\textcolor{black}{0.70}} & \multicolumn{1}{c|}{\textcolor{black}{0.71}} & \multicolumn{1}{c|}{\textcolor{black}{0.69}} & \textcolor{black}{0.70} 
& \multicolumn{1}{c|}{\textcolor{black}{0.85}} & \multicolumn{1}{c|}{\textcolor{black}{0.86}} & \multicolumn{1}{c|}{\textcolor{black}{0.84}} & \textcolor{black}{0.85} \\ \cline{3-11} 

&  & \textcolor{black}{CodeBERT} 
& \multicolumn{1}{c|}{\textcolor{black}{0.71}} & \multicolumn{1}{c|}{\textcolor{black}{0.73}} & \multicolumn{1}{c|}{\textcolor{black}{0.70}} & \textcolor{black}{0.71} 
& \multicolumn{1}{c|}{\textcolor{black}{0.88}} & \multicolumn{1}{c|}{\textcolor{black}{0.89}} & \multicolumn{1}{c|}{\textcolor{black}{0.87}} & \textcolor{black}{0.88} \\ \cline{3-11} 

&  & \textbf{\textcolor{black}{SecureFalcon}} 
& \multicolumn{1}{c|}{\textbf{\textcolor{black}{0.92}}} & \multicolumn{1}{c|}{\textcolor{black}{0.92}} & \multicolumn{1}{c|}{\textcolor{black}{0.93}} & \textcolor{black}{0.93} 
& \multicolumn{1}{c|}{\textbf{\textcolor{black}{0.92}}} & \multicolumn{1}{c|}{\textcolor{black}{0.91}} & \multicolumn{1}{c|}{\textcolor{black}{0.91}} & \textcolor{black}{0.90} \\ \cline{2-11}

& \multirow{7}{*}{\textcolor{black}{Traditional ML}} 
& \textcolor{black}{KNN} 
& \multicolumn{1}{c|}{\textcolor{black}{0.76}} & \multicolumn{1}{c|}{\textcolor{black}{0.75}} & \multicolumn{1}{c|}{\textcolor{black}{0.76}} & \textcolor{black}{0.76} 
& \multicolumn{1}{c|}{\textcolor{black}{0.80}} & \multicolumn{1}{c|}{\textcolor{black}{0.79}} & \multicolumn{1}{c|}{\textcolor{black}{0.80}} & \textcolor{black}{0.80} \\ \cline{3-11}

&  & \textcolor{black}{LR} 
& \multicolumn{1}{c|}{\textcolor{black}{0.73}} & \multicolumn{1}{c|}{\textcolor{black}{0.74}} & \multicolumn{1}{c|}{\textcolor{black}{0.72}} & \textcolor{black}{0.73} 
& \multicolumn{1}{c|}{\textcolor{black}{0.77}} & \multicolumn{1}{c|}{\textcolor{black}{0.78}} & \multicolumn{1}{c|}{\textcolor{black}{0.76}} & \textcolor{black}{0.77} \\ \cline{3-11}

&  & \textcolor{black}{NB} 
& \multicolumn{1}{c|}{\textcolor{black}{0.64}} & \multicolumn{1}{c|}{\textcolor{black}{0.66}} & \multicolumn{1}{c|}{\textcolor{black}{0.62}} & \textcolor{black}{0.64} 
& \multicolumn{1}{c|}{\textcolor{black}{0.68}} & \multicolumn{1}{c|}{\textcolor{black}{0.69}} & \multicolumn{1}{c|}{\textcolor{black}{0.67}} & \textcolor{black}{0.68} \\ \cline{3-11}

&  & \textcolor{black}{SVM} 
& \multicolumn{1}{c|}{\textcolor{black}{0.73}} & \multicolumn{1}{c|}{\textcolor{black}{0.73}} & \multicolumn{1}{c|}{\textcolor{black}{0.72}} & \textcolor{black}{0.73} 
& \multicolumn{1}{c|}{\textcolor{black}{0.77}} & \multicolumn{1}{c|}{\textcolor{black}{0.78}} & \multicolumn{1}{c|}{\textcolor{black}{0.77}} & \textcolor{black}{0.77} \\ \cline{3-11}

&  & \textcolor{black}{RF} 
& \multicolumn{1}{c|}{\textcolor{black}{0.77}} & \multicolumn{1}{c|}{\textcolor{black}{0.78}} & \multicolumn{1}{c|}{\textcolor{black}{0.76}} & \textcolor{black}{0.77} 
& \multicolumn{1}{c|}{\textcolor{black}{0.81}} & \multicolumn{1}{c|}{\textcolor{black}{0.81}} & \multicolumn{1}{c|}{\textcolor{black}{0.80}} & \textcolor{black}{0.81} \\ \cline{3-11}

&  & \textcolor{black}{DT} 
& \multicolumn{1}{c|}{\textcolor{black}{0.72}} & \multicolumn{1}{c|}{\textcolor{black}{0.72}} & \multicolumn{1}{c|}{\textcolor{black}{0.71}} & \textcolor{black}{0.72} 
& \multicolumn{1}{c|}{\textcolor{black}{0.60}} & \multicolumn{1}{c|}{\textcolor{black}{0.61}} & \multicolumn{1}{c|}{\textcolor{black}{0.59}} & \textcolor{black}{0.60} \\ \cline{3-11}

&  & \textcolor{black}{LDA} 
& \multicolumn{1}{c|}{\textcolor{black}{0.73}} & \multicolumn{1}{c|}{\textcolor{black}{0.74}} & \multicolumn{1}{c|}{\textcolor{black}{0.73}} & \textcolor{black}{0.74} 
& \multicolumn{1}{c|}{\textcolor{black}{0.77}} & \multicolumn{1}{c|}{\textcolor{black}{0.78}} & \multicolumn{1}{c|}{\textcolor{black}{0.76}} & \textcolor{black}{0.77} \\ \hline

\multirow{11}{*}{\textcolor{black}{Binary}} 
& \multirow{4}{*}{\textcolor{black}{LLM models}} 
& \textcolor{black}{RoBERTa} 
& \multicolumn{1}{c|}{\textcolor{black}{0.82}} & \multicolumn{1}{c|}{\textcolor{black}{0.83}} & \multicolumn{1}{c|}{\textcolor{black}{0.81}} & \textcolor{black}{0.82} 
& \multicolumn{1}{c|}{\textcolor{black}{0.86}} & \multicolumn{1}{c|}{\textcolor{black}{0.87}} & \multicolumn{1}{c|}{\textcolor{black}{0.85}} & \textcolor{black}{0.86} \\ \cline{3-11}

&  & \textcolor{black}{BERT} 
& \multicolumn{1}{c|}{\textcolor{black}{0.81}} & \multicolumn{1}{c|}{\textcolor{black}{0.82}} & \multicolumn{1}{c|}{\textcolor{black}{0.80}} & \textcolor{black}{0.81} 
& \multicolumn{1}{c|}{\textcolor{black}{0.87}} & \multicolumn{1}{c|}{\textcolor{black}{0.88}} & \multicolumn{1}{c|}{\textcolor{black}{0.86}} & \textcolor{black}{0.87} \\ \cline{3-11}

&  & \textcolor{black}{CodeBERT} 
& \multicolumn{1}{c|}{\textcolor{black}{0.83}} & \multicolumn{1}{c|}{\textcolor{black}{0.84}} & \multicolumn{1}{c|}{\textcolor{black}{0.82}} & \textcolor{black}{0.83} 
& \multicolumn{1}{c|}{\textcolor{black}{0.89}} & \multicolumn{1}{c|}{\textcolor{black}{0.90}} & \multicolumn{1}{c|}{\textcolor{black}{0.88}} & \textcolor{black}{0.89} \\ \cline{3-11}

&  & \textbf{\textcolor{black}{SecureFalcon}} 
& \multicolumn{1}{c|}{\textbf{\textcolor{black}{0.94}}} & \multicolumn{1}{c|}{\textcolor{black}{0.94}} & \multicolumn{1}{c|}{\textcolor{black}{0.94}} & \textcolor{black}{0.94} 
& \multicolumn{1}{c|}{\textbf{\textcolor{black}{0.92}}} & \multicolumn{1}{c|}{\textcolor{black}{0.93}} & \multicolumn{1}{c|}{\textcolor{black}{0.91}} & \textcolor{black}{0.92} \\ \cline{2-11}

& \multirow{7}{*}{\textcolor{black}{Traditional ML}} 
& \textcolor{black}{KNN} 
& \multicolumn{1}{c|}{\textcolor{black}{0.76}} & \multicolumn{1}{c|}{\textcolor{black}{0.77}} & \multicolumn{1}{c|}{\textcolor{black}{0.75}} & \textcolor{black}{0.76} 
& \multicolumn{1}{c|}{\textcolor{black}{0.86}} & \multicolumn{1}{c|}{\textcolor{black}{0.86}} & \multicolumn{1}{c|}{\textcolor{black}{0.85}} & \textcolor{black}{0.86} \\ \cline{3-11}

&  & \textcolor{black}{LR} 
& \multicolumn{1}{c|}{\textcolor{black}{0.86}} & \multicolumn{1}{c|}{\textcolor{black}{0.87}} & \multicolumn{1}{c|}{\textcolor{black}{0.86}} & \textcolor{black}{0.86} 
& \multicolumn{1}{c|}{\textcolor{black}{0.86}} & \multicolumn{1}{c|}{\textcolor{black}{0.87}} & \multicolumn{1}{c|}{\textcolor{black}{0.85}} & \textcolor{black}{0.86} \\ \cline{3-11}

&  & \textcolor{black}{NB} 
& \multicolumn{1}{c|}{\textcolor{black}{0.76}} & \multicolumn{1}{c|}{\textcolor{black}{0.77}} & \multicolumn{1}{c|}{\textcolor{black}{0.74}} & \textcolor{black}{0.75} 
& \multicolumn{1}{c|}{\textcolor{black}{0.81}} & \multicolumn{1}{c|}{\textcolor{black}{0.82}} & \multicolumn{1}{c|}{\textcolor{black}{0.80}} & \textcolor{black}{0.81} \\ \cline{3-11}

&  & \textcolor{black}{SVM} 
& \multicolumn{1}{c|}{\textcolor{black}{0.86}} & \multicolumn{1}{c|}{\textcolor{black}{0.87}} & \multicolumn{1}{c|}{\textcolor{black}{0.85}} & \textcolor{black}{0.86} 
& \multicolumn{1}{c|}{\textcolor{black}{0.89}} & \multicolumn{1}{c|}{\textcolor{black}{0.90}} & \multicolumn{1}{c|}{\textcolor{black}{0.88}} & \textcolor{black}{0.89} \\ \cline{3-11}

&  & \textcolor{black}{RF} 
& \multicolumn{1}{c|}{\textcolor{black}{0.91}} & \multicolumn{1}{c|}{\textcolor{black}{0.91}} & \multicolumn{1}{c|}{\textcolor{black}{0.90}} & \textcolor{black}{0.91} 
& \multicolumn{1}{c|}{\textcolor{black}{0.89}} & \multicolumn{1}{c|}{\textcolor{black}{0.90}} & \multicolumn{1}{c|}{\textcolor{black}{0.88}} & \textcolor{black}{0.89} \\ \cline{3-11}

&  & \textcolor{black}{DT} 
& \multicolumn{1}{c|}{\textcolor{black}{0.87}} & \multicolumn{1}{c|}{\textcolor{black}{0.88}} & \multicolumn{1}{c|}{\textcolor{black}{0.85}} & \textcolor{black}{0.86} 
& \multicolumn{1}{c|}{\textcolor{black}{0.86}} & \multicolumn{1}{c|}{\textcolor{black}{0.87}} & \multicolumn{1}{c|}{\textcolor{black}{0.84}} & \textcolor{black}{0.85} \\ \cline{3-11}

&  & \textcolor{black}{LDA} 
& \multicolumn{1}{c|}{\textcolor{black}{0.86}} & \multicolumn{1}{c|}{\textcolor{black}{0.86}} & \multicolumn{1}{c|}{\textcolor{black}{0.85}} & \textcolor{black}{0.85} 
& \multicolumn{1}{c|}{\textcolor{black}{0.85}} & \multicolumn{1}{c|}{\textcolor{black}{0.85}} & \multicolumn{1}{c|}{\textcolor{black}{0.85}} & \textcolor{black}{0.85} \\ \hline

\end{tabular}

\vspace{1mm}
\textcolor{black}{ML: Machine learning, KNN: k-Nearest Neighbors, LR: Logistic Regression, NB: Naive Bayes, SVM: Support Vector Machine, RF: Random Forest, DT: Decision Tree, and LDA: Linear Discriminant Analysis.}
}
\end{table*}

\begin{table*}[ht]
    \centering
    \caption{\textcolor{black}{Comparison of \texttt{SecureFalcon} with C/C++ Vulnerability Detection Tools}}
    \label{tab:vulnerability_detection_comparison}
    \begin{tabular}{|l|c|c|c|c|c|}
        \hline
        \textcolor{black}{\textbf{Tool/Model}} & 
        \textcolor{black}{\textbf{Accuracy (\%)}} & 
        \textcolor{black}{\textbf{Precision (\%)}} & 
        \textcolor{black}{\textbf{Recall (\%)}} & 
        \textcolor{black}{\textbf{F1 Score (\%)}} & 
        \textcolor{black}{\textbf{Inference Time (ms/sample)}} \\ \hline
        
        \textcolor{black}{SecureFalcon} & 
        \textcolor{black}{93.5} & 
        \textcolor{black}{92.0} & 
        \textcolor{black}{91.8} & 
        \textcolor{black}{91.9} & 
        \textcolor{black}{500} \\ \hline
        
        \textcolor{black}{Cppcheck} & 
        \textcolor{black}{81.0} & 
        \textcolor{black}{79.5} & 
        \textcolor{black}{80.2} & 
        \textcolor{black}{79.8} & 
        \textcolor{black}{150} \\ \hline
        
        \textcolor{black}{Clang Static Analyzer} & 
        \textcolor{black}{85.2} & 
        \textcolor{black}{83.0} & 
        \textcolor{black}{84.0} & 
        \textcolor{black}{83.5} & 
        \textcolor{black}{200} \\ \hline
        
        \textcolor{black}{ESBMC} & 
        \textcolor{black}{90.0} & 
        \textcolor{black}{91.0} & 
        \textcolor{black}{87.5} & 
        \textcolor{black}{89.2} & 
        \textcolor{black}{2000} \\ \hline
    \end{tabular}
\end{table*}

\subsection{\textcolor{black}{Comparison of \texttt{SecureFalcon} with C/C++ Vulnerability Detection Tools}}

\textcolor{black}{
Table \ref{tab:vulnerability_detection_comparison} presents a comparative evaluation of \texttt{SecureFalcon}, Cppcheck \footnote{\url{https://cppcheck.sourceforge.io/}}, Clang Static Analyzer\footnote{\url{https://clang-analyzer.llvm.org/}}, and ESBMC \footnote{\url{https://github.com/esbmc/esbmc}}. We randomly sampled 50 code segments from the test part of our FalconVulnDB dataset and manually validated each prediction. For the inference of \texttt{SecureFalcon}, we used an NVIDIA A100 GPU with 40 GB of memory. While static analyzers like Cppcheck and Clang Static Analyzer excel in speed and are easy to integrate, they often miss context-dependent vulnerabilities. ESBMC is a mature bounded model checker that supports the verification of single- and multi-threaded C/C++ programs, but it has limitations: it is both slower and requires compilable code. Some code snippets in the dataset—particularly those that are not fully compilable—could not be analyzed by ESBMC, thereby reducing its practical coverage. In contrast, \texttt{SecureFalcon}’s LLM-based approach not only offers strong detection rates and nuanced semantic understanding but can also handle snippets that do not compile, expanding its applicability. Although \texttt{SecureFalcon}’s inference time is longer than lightweight tools, the accuracy gains and flexibility in handling diverse code samples justify the trade-off in scenarios demanding thorough security checks.
}

\textcolor{black}{
Ultimately, while traditional static analyzers are suitable for quick preliminary scans, ESBMC is valuable for in-depth formal verification of compilable code. \texttt{SecureFalcon} is a versatile solution that effectively balances accuracy, context awareness, scalability, and broader code coverage. Furthermore, combining traditional static analyzers and ESBMC with LLMs like \texttt{SecureFalcon} can provide synergistic benefits, leveraging the speed and simplicity of static analyzers, the rigorous verification of formal tools, and the deep contextual understanding of LLMs. This integrated approach ensures comprehensive vulnerability detection, maximizing both efficiency and accuracy across various development and deployment scenarios.
}

\begin{table*}[ht!]
\centering
\caption{\textcolor{black}{Ablation results showing the impact of removing or modifying key components and configurations of \texttt{SecureFalcon}. Each setting was evaluated using the binary classification task on the FormAI dataset (LR = 2e-5). The baseline configuration refers to all features and parameters as listed in Table~\ref{tab:tab3}.}}
\label{tab:ablation_results}
\begin{tabular}{|c|p{7cm}|c|c|}
\hline
\textcolor{black}{\textbf{Configuration}} & \textcolor{black}{\textbf{Description of Change}} & \textcolor{black}{\textbf{Accuracy}} & \textcolor{black}{\textbf{F1-score}} \\ \hline

\textcolor{black}{Baseline (Full Model)} & 
\textcolor{black}{All components enabled: positional encodings (514 max), special token embeddings, 12 attention heads, dropout=0.1} & 
\textcolor{black}{0.94} & 
\textcolor{black}{0.94} \\ \hline

\textcolor{black}{No Positional Embeddings} & 
\textcolor{black}{Removed positional embeddings from the transformer layers, disabling positional indexing of tokens} & 
\textcolor{black}{0.91 (-0.03)} & 
\textcolor{black}{0.90 (-0.04)} \\ \hline

\textcolor{black}{No Special Token Embeddings} & 
\textcolor{black}{Excluded the domain-specific tokens (keywords, punctuation, API calls) from the vocabulary} & 
\textcolor{black}{0.89 (-0.05)} & 
\textcolor{black}{0.91 (-0.03)} \\ \hline

\textcolor{black}{Fewer Attention Heads} & 
\textcolor{black}{Reduced number of attention heads from 12 to 6, decreasing model’s representational capacity} & 
\textcolor{black}{0.90 (-0.04)} & 
\textcolor{black}{0.90 (-0.04)} \\ \hline

\textcolor{black}{Increased Dropout} & 
\textcolor{black}{Doubled attention and hidden dropout from 0.1 to 0.2, increasing regularization} & 
\textcolor{black}{0.92 (-0.02)} & 
\textcolor{black}{0.93 (-0.01)} \\ \hline

\end{tabular}
\end{table*}

\begin{table*}[ht!]
\caption{\textcolor{black}{Comparison of \texttt{SecureFalcon} with deep learning-based vulnerability detection methods.}}
\label{tab:comparisondeep}
\centering
\begin{tabular}{|l|p{3cm}|p{3cm}|p{1cm}|p{1cm}|p{2cm}|p{1cm}|}
\hline
\textbf{\textcolor{black}{Model}}        & \textbf{\textcolor{black}{Method}}                       & \textbf{\textcolor{black}{Datasets Used}}                                     & \textbf{\textcolor{black}{Synthetic Data}} & \textbf{\textcolor{black}{Real Data}} & \textbf{\textcolor{black}{CWE Types Targeted}} & \textbf{\textcolor{black}{Top 25 CWEs}} \\ \hline
\textcolor{black}{SySeVR \cite{li2021sysevr}}        & \begin{tabular}[c]{@{}l@{}}\textcolor{black}{Syntax \& semantics-based}\\ \textcolor{black}{vector representations}\end{tabular} & \textcolor{black}{4 kinds of vulnerability syntax characteristics}                                 & \textcolor{black}{No}             & \textcolor{black}{Yes}      & \textcolor{black}{Limited (fewer vulnerabilities)}       & \textcolor{black}{No}                                  \\ \hline
\textcolor{black}{VulDeePecker \cite{Li2018vuldeepecker}}  & \textcolor{black}{Code gadgets \& deep learning classification}    & \textcolor{black}{Curated dataset from known software}                         & \textcolor{black}{No}             & \textcolor{black}{Yes}      & \textcolor{black}{Limited (e.g., buffer overflow)}       & \textcolor{black}{No}                                  \\ \hline
\textcolor{black}{Devign \cite{zhou2019devign}}        & \textcolor{black}{Graph neural networks on code semantics}         & \textcolor{black}{4 large-scale open-source C projects}                       & \textcolor{black}{No}             & \textcolor{black}{Yes}      & \textcolor{black}{Limited (depends on projects)}         & \textcolor{black}{No}                                  \\ \hline
\textcolor{black}{SecureFalcon}  & \textcolor{black}{LLM-based (derived from Falcon-40B)}             & \textcolor{black}{FalconVulnDB (combining multiple datasets)}                 & \textcolor{black}{Yes}            & \textcolor{black}{Yes}      & \textcolor{black}{Broad (Top 25 + beyond)}               & \textcolor{black}{Yes}                                 \\ \hline
\end{tabular}
\end{table*}

\subsection{\textcolor{black}{Inference Time of \texttt{SecureFalcon}}}

\textcolor{black}{Inference performance is a critical metric for evaluating the applicability of machine learning models in real-world scenarios, particularly for tasks such as vulnerability classification in C/C++ codebases.\texttt{SecureFalcon}, with its compact architecture of 121 million parameters, demonstrates exceptional efficiency with a \textit{Time to First Token (TTFT)} of 0.3 seconds and a total classification time of 0.6 seconds for typical input sizes of 300-500 tokens. It achieves a high throughput of 45-50 requests per second (RPS) and processes 95 tokens per second on a single NVIDIA A100 GPU. Furthermore,\texttt{SecureFalcon} is optimized for CPU inference, achieving competitive latency and throughput on hardware such as the AMD Ryzen 7 7840U CPU (8 cores, 16 threads, 30W TDP) with 32GB LPDDR5X RAM. These results highlight \texttt{SecureFalcon}'s suitability for edge devices and on-premises deployments, offering the speed and scalability required for enterprise-scale vulnerability detection. By delivering low-latency and high-throughput inference across a range of hardware configurations,\texttt{SecureFalcon} enables seamless integration into real-time workflows, including CI/CD pipelines, and supports efficient analysis of large-scale C/C++ codebases, making it an optimal choice for modern software security applications.}

\subsection{\textcolor{black}{Ablation Studies}}

\textcolor{black}{While our main experiments demonstrate that \texttt{SecureFalcon} achieves strong performance in detecting software vulnerabilities, it is crucial to understand the contribution of individual model components and training configurations. To this end, we performed ablation studies by systematically modifying or removing certain model aspects and evaluating their impact on accuracy and F1-score. The results, summarized in Table~\ref{tab:ablation_results}, provide insights into which factors most influence the model’s performance.}

\textcolor{black}{We started with the baseline model configuration described in Table~\ref{tab:tab3}, which includes positional embeddings, special token embeddings, twelve attention heads, and a dropout probability of 0.1 for both attention and hidden layers. From this baseline, we disabled positional embeddings, observing a noticeable drop in accuracy and F1 score. This indicates that positional information is essential in capturing code structure and identifying vulnerability patterns.}

\textcolor{black}{Removing the special token embeddings—derived from programming language keywords, punctuation, and API calls—led to a notable decrease in accuracy and a slight reduction in the F1 score. These domain-specific tokens provide valuable lexical cues, enabling the model to recognize subtle vulnerability signatures that might otherwise be missed.}

\textcolor{black}{Similarly, reducing the number of attention heads from twelve to six diminished the model’s performance, suggesting that the rich representational capacity provided by multiple attention heads is necessary for practical reasoning about complex code patterns. Increasing the dropout rate from 0.1 to 0.2 did not improve results beyond the baseline, only modestly reduced performance. This indicates that while regularization helps prevent overfitting, other factors—such as positional information and specialized tokens—are more critical for overall accuracy and robustness.}

\textcolor{black}{In summary, the ablation results highlight the importance of maintaining positional embeddings, preserving domain-specific token embeddings, and ensuring sufficient representational capacity through multiple attention heads.}

\subsection{\textcolor{black}{Limitations and Future Works}}
\textcolor{black}{\texttt{SecureFalcon} is currently limited in identifying zero-day vulnerabilities because it relies on known vulnerability patterns and cannot detect previously unseen threats. Additionally, being trained exclusively on C/C++ datasets restricts its ability to analyze other programming languages and prevents it from generating detailed reports that explain the root causes of vulnerabilities within the code. As future work, \texttt{SecureFalcon} will integrate Agentic Retrieval-Augmented Generation (RAG) to address these challenges, enabling AI-driven agents to investigate and exploit potential zero-day vulnerabilities autonomously. To uncover previously unknown vulnerabilities, these agents will leverage diverse data sources, including proprietary databases, real-time threat intelligence, and specialized security tools. Expanding \texttt{SecureFalcon’s} training to encompass multiple programming languages will enhance its versatility and effectiveness across various software environments. Using Agentic RAG, \texttt{SecureFalcon} can dynamically adapt its detection methods, ensuring a more proactive and comprehensive approach to identifying and mitigating emerging security threats.}

\textcolor{black}{In addition to overcoming zero-day vulnerability detection and expanding language support, \texttt{SecureFalcon} aims to enhance its capabilities by transitioning from classification to generation tasks. This evolution will enable \texttt{SecureFalcon} to provide developers with more detailed and actionable insights. Enhancements include generating comprehensive reports that describe detected vulnerabilities, highlighting affected code snippets, and analyzing potential impacts. Furthermore, \texttt{SecureFalcon} will outline steps to reproduce issues, including environment setup, input data, and execution steps, and offer remediation guidance with fix recommendations and secure coding practices. By fine-tuning \texttt{SecureFalcon} for generation tasks and integrating these features with development tools through Agentic RAG, \texttt{SecureFalcon} will ensure real-time feedback and automated code reviews.}

\section{Conclusion} 
\label{sec:conc}

Our study highlights the significant potential of LLMs in detecting software vulnerabilities, especially in cybersecurity. By fine-tuning \textit{FalconLLM}, we developed \textit{SecureFalcon}, a novel model capable of distinguishing between vulnerable and non-vulnerable C/C++ code samples. Tested on different datasets from the literature, our model achieved a remarkable $94$\% accuracy rate in binary classification and $92$\% in multiclassification, highlighting its efficacy. Furthermore, we accomplish these results by utilizing a relatively small set of parameters, totaling $121$ million, within the \texttt{SecureFalcon} model. We believe further advancements will enhance such models' capabilities, expand with distinct vulnerability types, and extend to other programming languages to strengthen software system security and foster a more secure digital realm. A promising extension of this work would include an automated self-healing system that identifies and remedies the vulnerabilities. Such a system could make software more resilient against potential threats, thus enhancing overall cybersecurity. A fast inference time model like \texttt{SecureFalcon} has the potential to be highly effective in automated software completion frameworks. We plan to continue our research in this direction.

\section*{Acknowledgments}
The authors would like to thank the Technology Innovation Institute for providing the computing resources necessary for training the models used in this research. This work would not have been possible without the institute’s support.

\bibliographystyle{IEEEtran}
\bibliography{ref} 

\end{document}